\documentclass[fleqn,usenatbib]{mnras}

\usepackage{newtxtext,newtxmath}

\usepackage[T1]{fontenc}
\usepackage{ae,aecompl}

\usepackage{float}
\usepackage[caption=false]{subfig}
\usepackage{mathtools}
\usepackage{mwe}
\usepackage{graphicx}	
\usepackage{amsmath}	
\usepackage{amssymb}	

\title[332~MHz drift scan survey]{Limits on Absorption from a 332-MHz survey for Fast Radio Bursts}

\author[K. M. Rajwade et al.]{
K. M. Rajwade,$^{1}$\thanks{E-mail: kaustubh.rajwade@manchester.ac.uk}
M. B. Mickaliger,$^{1}$
B. W. Stappers,$^{1}$
C. G. Bassa,$^{2}$
\newauthor
R. P. Breton,$^{1}$
A. Karastergiou$^{3}$
 and E. F. Keane$^{4,1}$
\\
$^{1}$Jodrell Bank Centre for Astrophysics, University of Manchester, Oxford Road, Manchester M13 9PL, UK\\
$^{2}$ASTRON, the Netherlands Institute for Radio Astronomy, Oude Hoogeveensedijk 4, 7991 PD
  Dwingeloo, The Netherlands \\
$^{3}$Astrophysics, Denys Wilkinson Building, University of Oxford, Keble road, Oxford OX1 3RH, UK\\
$^{4}$SKA Organisation, Jodrell Bank Observatory, Macclesfield SK11 9DL, UK
}

\date{Accepted XXX. Received YYY; in original form ZZZ}

\pubyear{2015}

\begin{document}
\label{firstpage}
\pagerange{\pageref{firstpage}--\pageref{lastpage}}
\maketitle

\begin{abstract}
Fast Radio Bursts (FRBs) are bright, extragalactic radio pulses whose origins are still unknown. Until recently, most FRBs have been detected at frequencies greater than 1~GHz with a few exceptions at 800~MHz. The recent discoveries of FRBs at 400~MHz from the Canadian Hydrogen Intensity Mapping Experiment (CHIME) telescope has opened up possibilities for new insights about the progenitors while many other low frequency surveys in the past have failed to find any FRBs. Here, we present results from a FRB survey recently conducted at the Jodrell Bank Observatory at 332~MHz with the 76-m Lovell telescope for a total of 58 days. We did not detect any FRBs in the survey and report a 90$\%$ upper limit of 5500 FRBs per day per sky for a Euclidean Universe above a fluence threshold of 46~Jy~ms. We discuss the possibility of absorption as the main cause of non-detections in low frequency (< 800~MHz) searches and invoke different absorption models to explain the same. We find that Induced Compton Scattering alone cannot account for absorption of radio emission and that our simulations favour a combination of Induced Compton Scattering and Free-Free Absorption to explain the non-detections. For a free-free absorption scenario, our constraints on the electron density are consistent with those expected in the post-shock region of the ionized ejecta in Super-Luminous SuperNovae (SLSNe).  


\end{abstract}



\section{Introduction}
FRBs are millisecond duration, bright radio signals that are observed over the entire sky. Based on their measured dispersion measures (DMs), the integrated electron density along the line of sight, they are extragalactic in origin. Since their discovery in 2007~\citep{lo07}, more than 64 have been published~\citep[see e.g.][]{th13,ch16,CHIME2018,sh2018}. The first FRB seen to repeat was FRB 121102~\citep{sp16}, leading to its localisation and the identification of the host~\citep{sp16,cha17}. This discovery has led to large all-sky surveys looking for FRBs and possibly repeating bursts from the same source leading to more discoveries and localisations~\citep{CHIME2019c,ban2019,ravi2019}. Most FRBs, including the first repeating FRB have been observed at 1.4~GHz and higher frequencies. Recently, the Canadian Hydrogen Intensity Mapping Experiment (CHIME) telescope detected 13 FRBs in the range of 400--800~MHz within a span of 10 weeks, significantly increasing the sample size of these sources~\citep{CHIME2018}.

Over the years, multiple low radio frequency ( < 800~MHz) searches have been conducted to search for FRBs. Despite many detections at 1.4~GHz, none have been seen at frequencies lower than 400~MHz.~\cite{coe2014} performed a large sky survey at 142~MHz using the LOw Frequency ARray (LOFAR)~\citep{vanHaarlem2013} and reported an upper limit on the rate of 150 events per day per sky above a flux of 107~Jy. A very stringent limit on the FRB rate at 145~MHz of 29 events per day per sky was reported by~\cite{ka15} using the Chilbolton, UK LOFAR station (Rawlings array) as they covered 4193 sq.deg of the sky in 60 days and did not find any FRBs. Non-detections in these surveys did hint at the possibility that FRB event rates at low frequencies are smaller compared to those reported at higher frequencies~\citep{sh2018, pe15}. This possibility was also supported by non-detections from extremely sensitive searches for FRBs conducted at 350~MHz using the Green Bank Telescope (GBT) and Arecibo observatory (AO)~\citep{ch17,de16} with the most constraining upper limit of 4980 events per day per sky at these frequencies for a flux limit of 0.61~Jy~\citep{ch17}. On the other hand, before CHIME became operational, the UTMOST survey~\citep{ca17} was successful in finding FRBs at a frequency of 843~MHz showing that FRBs did emit at frequencies below 1~GHz.  

The detection results from various surveys at different frequencies can be interpreted in many ways.~\cite{sh2018} report frequency structure in the detected bursts at 1.4~GHz. A significant portion of the flux of the burst lies within a very narrow band. This non-contiguous broadband behaviour can significantly affect the detection statistics of surveys with large bandwidths. Recently, the repeating FRB, FRB 121102 has been shown to exhibit variable spectral behaviour where the emission drifts across the frequency band.~\cite{hes2019} conducted a multi-frequency study of this behaviour and have shown that the drift rate increases at lower frequencies. This suggests that drifting sub-pulses across the frequency band, even in the complete absence of scattering, can cause the resulting pulse to be wider, thus decreasing the signal-to-noise ratio of the pulse and rendering them non-detectable at lower frequencies.

This dearth of low-frequency events has led a few authors to claim that absorption and scattering around the source play a significant role in mitigating radio emission at these frequencies~\citep{ku15,ravi2018a}.~\cite{sok2018} provided indirect evidence for this when the Murchison Wide-field Array (MWA) was shadowing part of the sky where seven FRBs were detected by ASKAP~\citep{sh2018} and none were detected at MWA frequencies. Based on the non-detection, they were able to obtain a lower limit on the optical depth of the plasma surrounding the progenitor assuming free-free absorption as the main cause. Currently, the consensus on whether absorption is a primary cause of non-detections is still an open question. Detecting FRBs below 400~MHz will be crucial for solving this puzzle.

Here, we present results from a drift scan search for FRBs at 332~MHz (92~cm) using the 76-m Lovell Telescope at Jodrell Bank, UK, during its scheduled summer maintenance. The frequency range of this survey was suitably placed such that it was just below the CHIME band but higher than the previous LOFAR searches and thus, any detection would go a long way in revealing valuable information about FRB emission at these frequencies. In Section \ref{section:survey}, we detail the survey and the FRB search pipeline, and discuss Radio Frequency Interference (RFI) mitigation techniques used. The results from the search and corresponding rate calculations are presented in Section \ref{section:results}. We present the results of our Monte-Carlo (MC) simulations to constrain potential absorption models in Section \ref{sec:MC}. Finally, we discuss our findings and conclude in Section \ref{sec:discussions}.

\section{The survey}
\label{section:survey}
\subsection{Survey Description}

The 332~MHz receiver was installed in the focus cabin of the Lovell Telescope in the spring of 2016. We performed a drift scan survey to search for FRBs during the summer months while the telescope was undergoing maintenance on the azimuthal track and pointed at the zenith. To achieve this, we used the ROACH backend~\citep{bassa2016} to coarsely channelize the band from 302 to 366\,MHz into 4 subbands of 16\,MHz. Each of these subbands was subsequently channelized to $32\times0.5$\,MHz channels, and time averaged to 256\,$\upmu$s samples using \texttt{digifil} from the \texttt{dspsr} software suite~\citep{vanStraten2012}. The finely channelized data of these four subbands were sent to another machine where the bands were stacked in frequency. Data were saved in 60-second chunks, which was chosen to balance the number and size of files.

\subsection{RFI Excision}
\label{section:RFI}

\begin{figure*}
\centering
\includegraphics[width=\columnwidth]{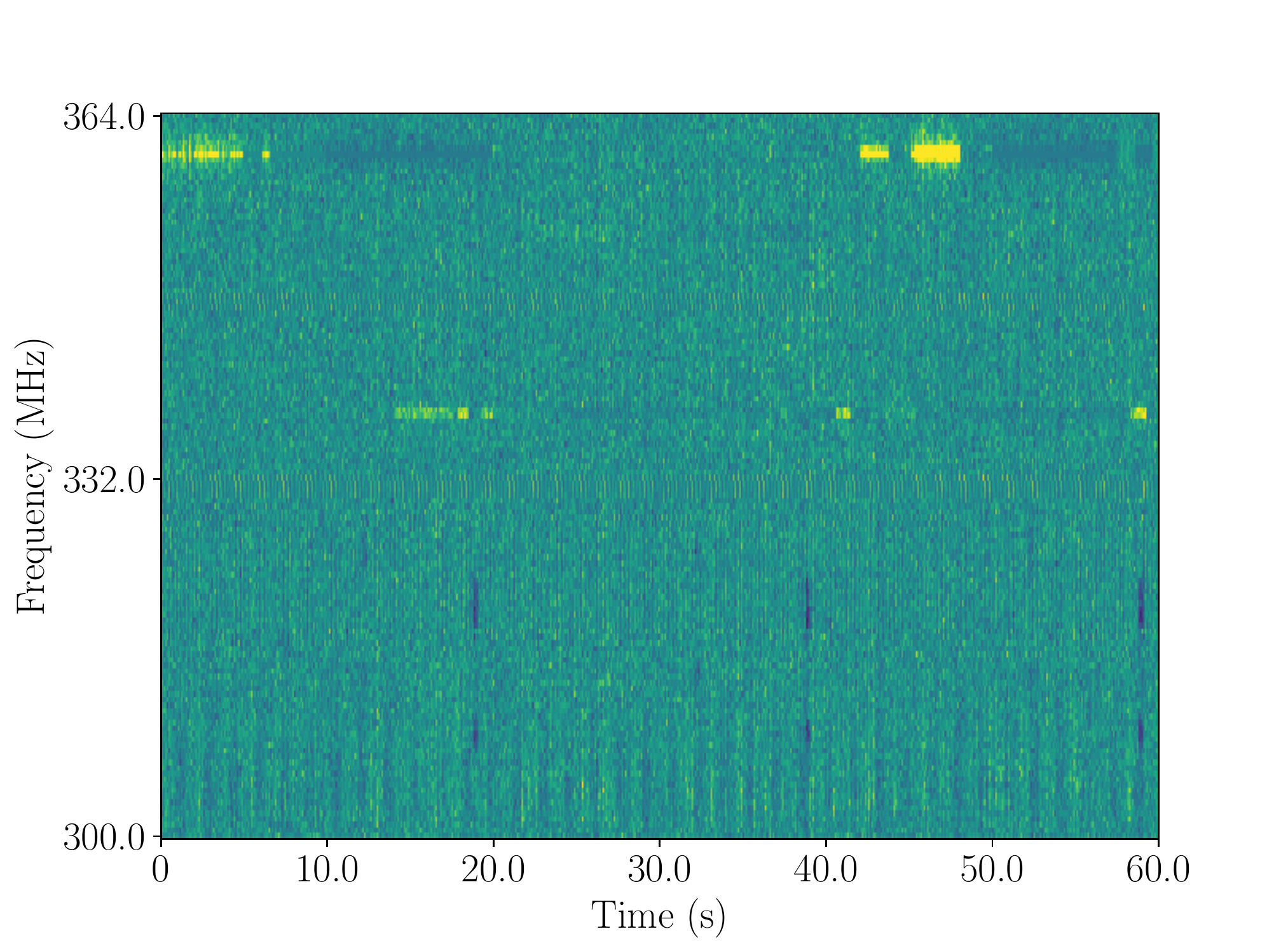} 
\includegraphics[width=\columnwidth]{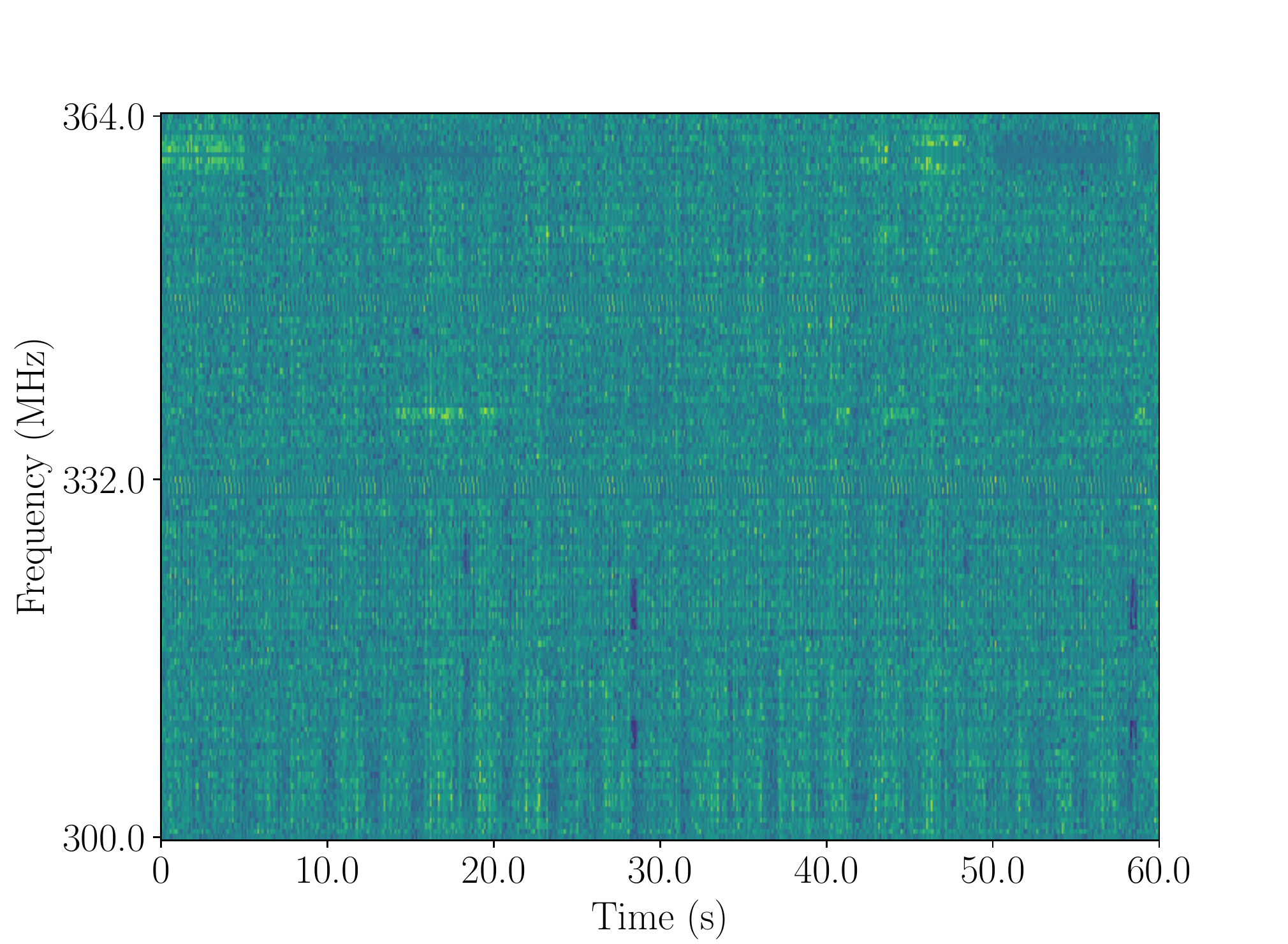} 
\caption{\textbf{Left Panel:} Raw dynamic spectrum of a 60-second chunk of our 332~MHz data on MJD 57870. \textbf{Right Panel:} Same chunk after running the MAD filter on the data.}
\label{fig:rfi}
\end{figure*}

\begin{figure}
\centering
\includegraphics[width=\columnwidth]{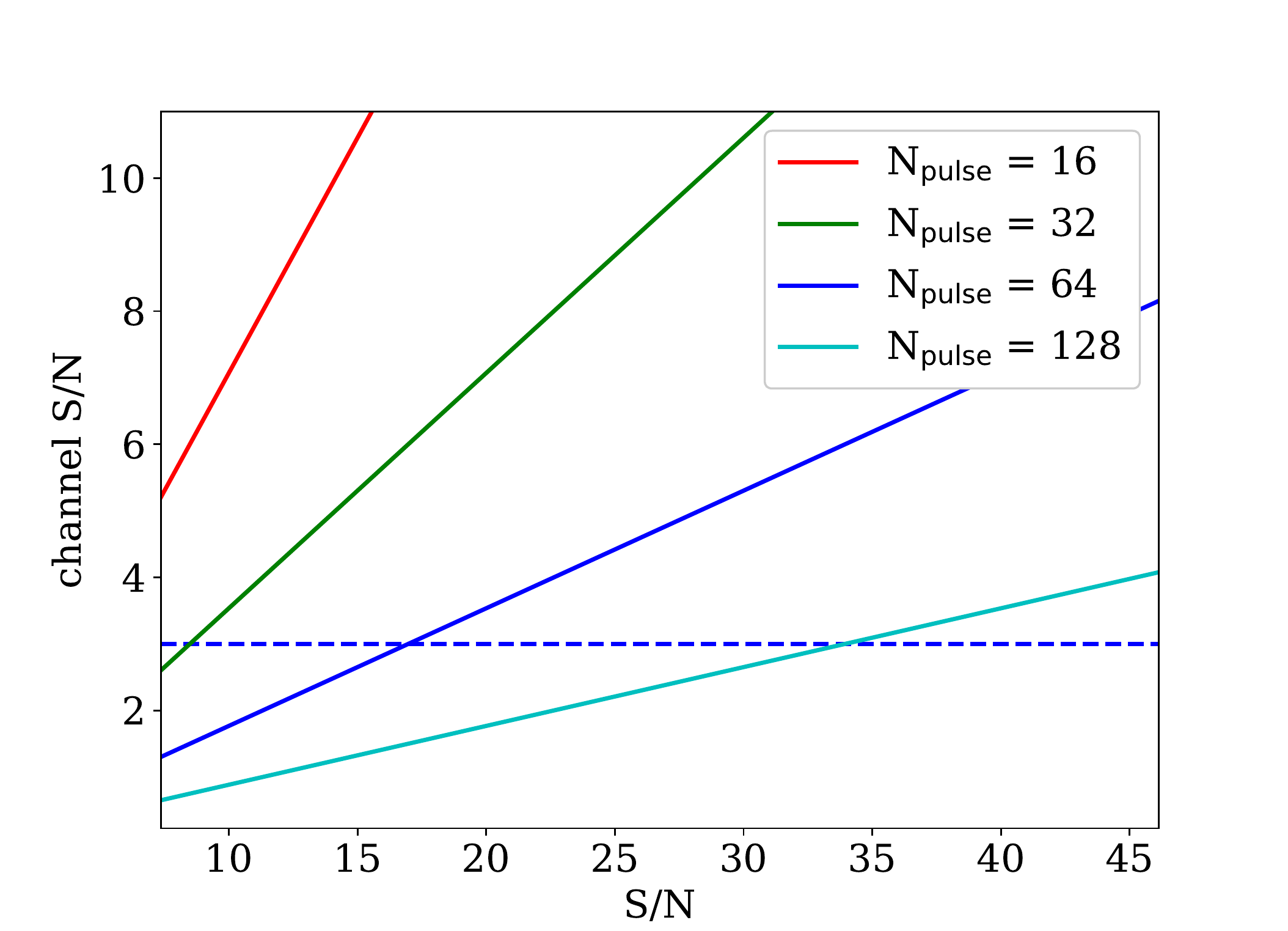}
\caption{Channel signal-to-noise ratio as a function of the total signal-to-noise ratio of a putative bright pulse for different frequency coverage of the signals. N$_{\rm pulse}$ denotes the number of channels over which the signal spans. The blue dashed horizontal line corresponds to the MAD threshold above which the pulses will be flagged as RFI.}
\label{fig:MAD}
\end{figure}

As the observations were in the summer maintenance period, the various construction equipment caused severe RFI. Moreover, the dish was pointing towards zenith which exposed the focus cabin to the horizon. Therefore, we had to implement RFI excision techniques before the data were rendered usable for processing. We first masked frequency channels that were always corrupted by RFI which reduced the total usable bandwidth by 10$\%$. Then, we filtered the dynamic spectrum using a Graphical Processing Unit (GPU)-based Median Absolute Deviation (MAD)~\citep{pr89} algorithm. In statistics, the median absolute deviation (MAD) is a robust measure of the variability of a univariate sample of quantitative data. For a univariate dataset $X_{1}$, $X_{2}$, $X_{3}$,...,$X_{n}$,
\begin{equation}
{\rm MAD} = {\rm median}\left(X_{i} - \tilde{X}\right),
\end{equation}
where $\tilde{X}$ is the median of the dataset. The MAD is related to the standard deviation by a scaling factor such that,
\begin{equation}
\sigma = \hat{k}\rm{MAD}.
\end{equation}
For a normally distributed dataset, $\hat{k}=1.4826$~\citep{Rousseeuw1993}.
Since MAD is a robust estimator of the noise statistics and is unaffected by narrow-band RFI, it can be used effectively to remove narrow-band bursts of RFI. Figure~\ref{fig:rfi} shows the effects of MAD filter on our dataset where one can clearly see the narrow-band RFI being flagged by the algorithm.

The MAD filter removed almost all the narrow-band RFI in the data. We used the method to clean data along the frequency axis as time-domain filtering will affect our signal of interest i.e. bright single pulses would also be flagged by the filtering algorithm in the time domain. This has implications on FRBs that are confined to a subset of channels in the band like the ones discovered by~\cite{sh2018}. We note that the ideal way to implement the MAD algorithm is after dedispersion has been performed on the dynamic spectrum and before frequency channels are added to make DM trials, since any bright pulse would be aligned in frequency and would not be affected. However, the gain in sensitivity due to RFI removal outweighed this issue.

In the MAD implementation, each GPU thread runs a MAD filter on all channels in one time sample. For a given astrophysical signal in a time sample, the signal-to-noise ratio of the signal per channel,
\begin{equation}
\mathrm{S/N_{\rm ts}} = \frac{\mathrm{S/N_{\rm  pulse}}}{\sqrt{N_{\rm chans}}},
\end{equation}
where S/N$_{\rm pulse}$ is the signal-to-noise ratio of the pulse after integrating over the full band and $N_{\rm chans}$ is the number of channels in the spectrum. We note that this assumes that the intrinsic frequency spectrum of the signal is flat over the observing band. We used a flagging threshold ($\rm th_{\rm MAD}$) of 3, so assuming Gaussian noise, the threshold in terms of the standard deviation of the noise,
\begin{equation}
\rm th_{\rm MAD} = \rm 3.0~\sigma =  4.44~MAD .
\end{equation}
Figure \ref{fig:MAD} shows the maximum S/N of the bright pulse before it is flagged as RFI by our algorithm for different spans of the FRB within the bandwidth of the survey. We note that our algorithm will flag astrophysical sources with S/N < 10 as RFI if the signal spans not more than a quarter of the band while a broadband signal will have to be extremely bright to be flagged as RFI. We computed the probability of finding a bright pulse (S/N > 30) that spans the entire band based on the best constraints on the log(N)--log(F) slope of FRBs~\citep{mac2018}, -2.8, and we find that the the probability of detecting a bright pulse with S/N > 30 above our detection threshold is $\sim$3$\times$10$^{-4}$. Nevertheless, we acknowledge the caveat that our pipeline will be insensitive to both, extremely bright and extremely narrow astrophysical signals.

\begin{figure}
\centering
\includegraphics[scale=0.65]{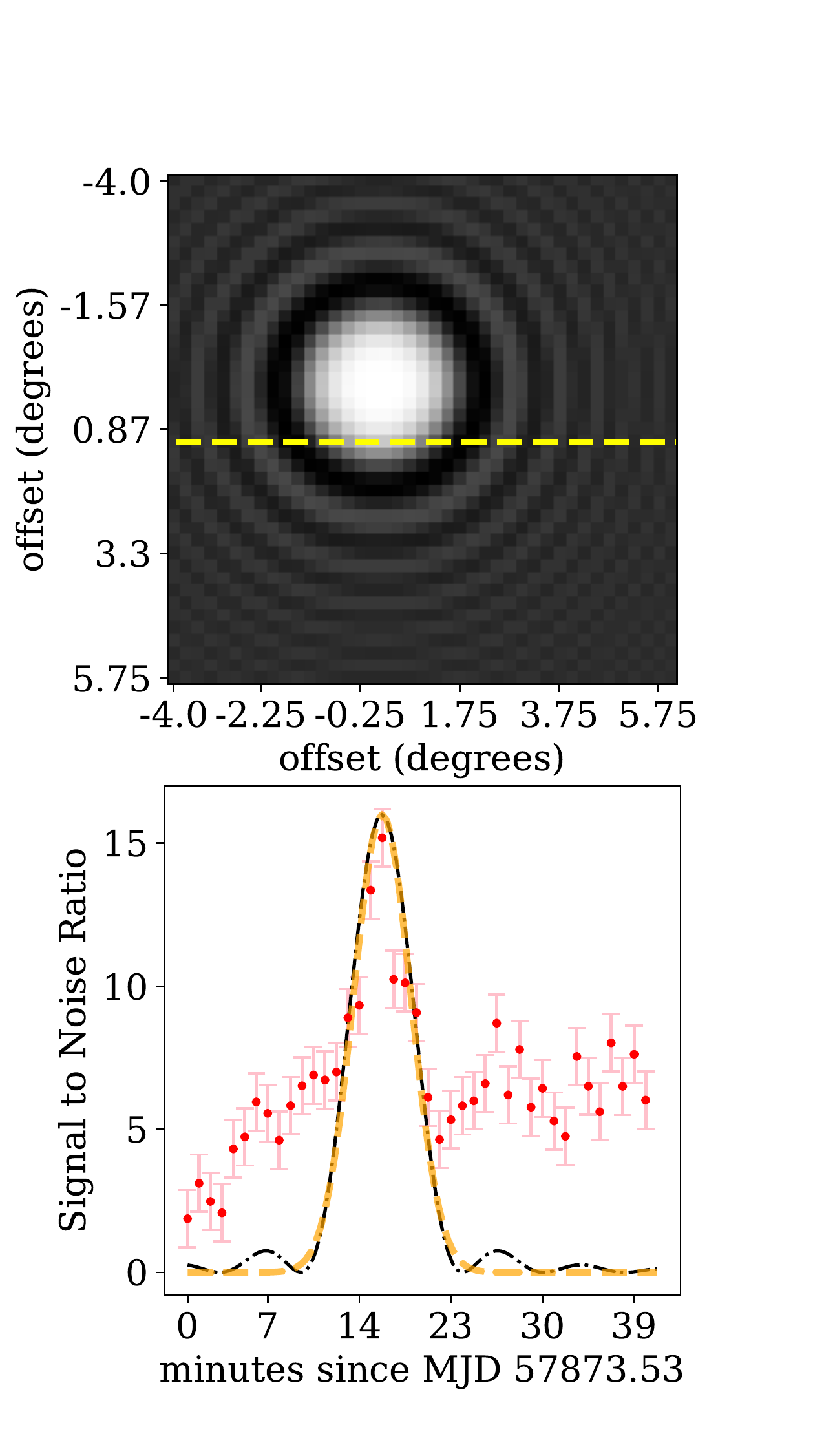}
\caption{\textbf{Top Panel}: Ideal Beam pattern of the 332~MHz receiver. The yellow horizontal line corresponds to the traverse of PSR~B0329+54 through the beam. \textbf{Bottom Panel}: S/N (red points) as a function of time for PSR B0329+54 as it drifts through the beam of the 332~MHz receiver. The black dot-dashed line denotes the beam shape of an ideal feed scaled to the maximum S/N of PSR B0329+54 in the beam along with the approximated Gaussian beam used in our analysis (orange dashed line). The receiver was stationary during the entire observation. The telescope was pointing at an azimuth of 170.001$^{\circ}$ and elevation of 88.83~$^{\circ}$. }
\label{fig:beam}
\end{figure}

\subsection{Sensitivity}
 We estimated the sensitivity of the search to potential FRBs at 332~MHz using the radiometer equation~\citep{lo04}. Since we do not know the position of the FRB in the beam, we compute the fluence threshold for our survey weighted by the beam pattern of the 332~MHz receiver. Thus, given a minimum detection threshold, S/N$_{\rm lim}$, the fluence limit,

\begin{equation}
S_{\rm peak}  =  \frac{\mathrm{{S/N}_{\rm lim}}~T_{\rm sys}}{G_{\rm 0} \sqrt{W_{\rm eff} n_{p}\Delta \nu}}\, \, \, \, \rm Jy ,
\label{eq:flux}
\end{equation}
where the position weighted gain,
\begin{equation}
G_{\rm 0} = \frac{\int_{0}^{\Omega_{\rm FoV}} G(\Omega) d\Omega }{\int_{0}^{\Omega_{\rm FoV}}  d\Omega}.
\end{equation}
Here, $\Omega_{\rm FoV}$ is the solid angle  subtended by the beam, $T_{\rm sys}$ is the system temperature, $G(\Omega)$ is position dependent gain, $W_{\rm eff}$ is the effective width of the pulse, $n_{\rm p}$ is number of polarisations that are summed, and $\Delta\nu$ is the bandwidth of the system.

\begin{figure*}
\centering
\includegraphics[scale=0.5]{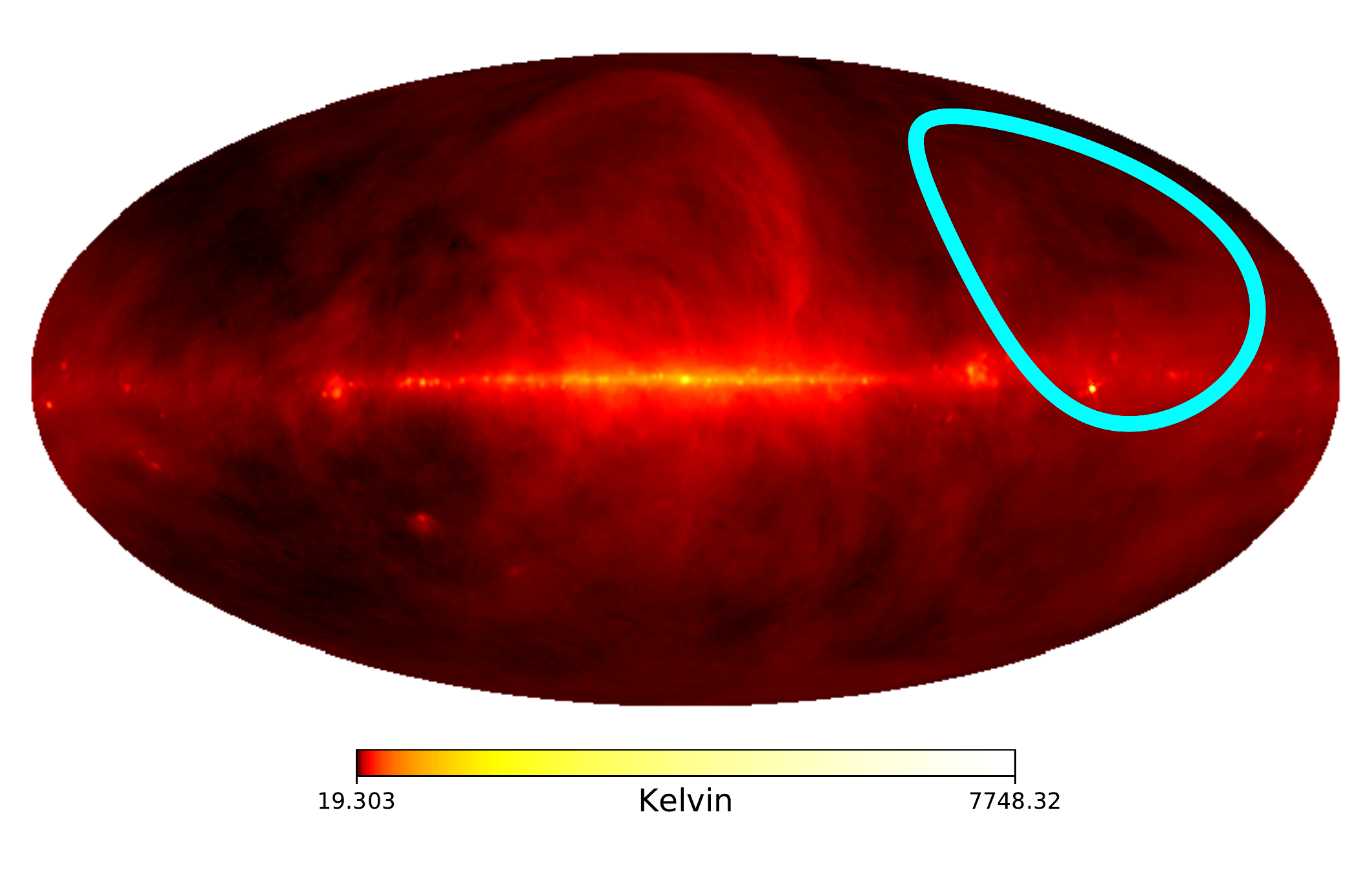}\\
\includegraphics[scale=0.6]{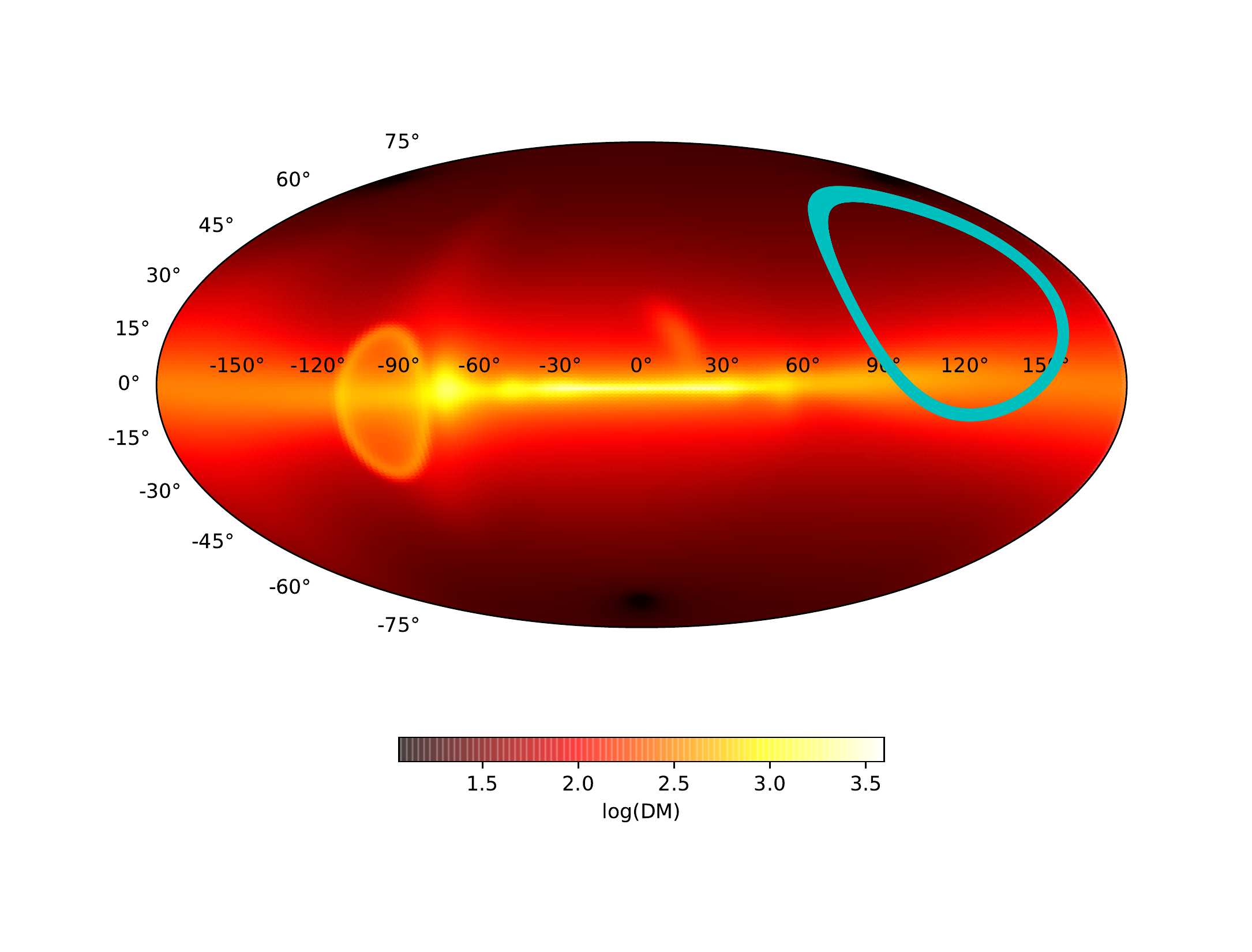}
\caption{All-sky map of the contribution of the Galactic synchrotron emission at 332~MHz (Top Panel) and the maximum DM contribution from the Galaxy using the YMW16 model of electron density in the Milky Way~\protect{\citep{Yao2017}} (Bottom Panel). The sky temperature has been scaled to 332~MHz from the 408~MHz all-sky map~\protect{\citep{rema2015}} with a spectral index of $-$2.5. The value of $-$2.5 is the mean value from the range of values reported in the literature~\protect{\citep[for e.g.][]{Guzmn2010, Jones2001}}. The cyan shaded region shows the surveyed region of sky.}
\label{fig:dm}
\end{figure*}

To measure the beam shape for the 332~MHz receiver, we decided to use a bright pulsar, namely PSR B0329+54 that fortuitously drifted through the primary beam and the first 2 sidelobes in which it was detectable throughout. We split our data into 1 minute chunks and folded them modulo the pulsar period. The measurement of S/N of each folded profile can be used to measure the response of the beam as a function of offset from the boresight (Figure~\ref{fig:beam}). Though the pulsar does not traverse through the boresight of the primary beam, the traverse does give us the beam response along the azimuthal axis. Although the primary beam of the 332~MHz receiver is symmetric and can be approximated by a Gaussian, the side-lobes are difficult to model analytically. This is evident in the shape of the S/N as seen in Figure.~\ref{fig:beam}. Moreover, the response of the sidelobes varied over different days that made it difficult to measure a beam pattern. Hence, we decided to compute the weighted gain upto the full width half max 
(FWHM) of the primary beam by modelling the beam as a Gaussian. Hence, the position dependent beam response,
\begin{equation}
G(\Omega) = G(\theta, \phi) = G~e^{\frac{-(\theta)^{2}}{2\sigma^{2}}},
\end{equation}
where $G$ is the gain at the boresight that is measured to be $\sim$0.9~K~Jy$^{-1}$ and $\sigma = 0.43~\rm FWHM$.
We assume the beam is axisymmetric up to the first null of the primary beam. Henceforth, we only use the primary beam up to the FWHM for all our rate calculations.
We assume that the the $T_{\rm sys}$ and $G(\Omega)$ do not vary with frequency, given our small bandwidth. Based on mulitple on-sky measurements at an elevation of 40$^\circ$, we used the measured $T_{\rm sys}$ = 350~K. We note that the $T_{\rm sys}$ measurements were done in a similar RFI environment so the measured value already includes average contribution from external RFI. We added an additional value of 50~K to this measured value to account for the average contribution from the Galactic synchrotron emission from the sky over the span of one full day (see top panel of figure~\ref{fig:dm}). From these values, we obtain a peak flux of 4.59~Jy for a 10~ms burst with a S/N of 9. This corresponds to a fluence limit of 45.9 Jy~ms at the boresight of the beam.

\subsection{Single Pulse search pipeline}
After filtering the RFI, we processed the data using the GPU-based single pulse search software suite called \textsc{heimdall}~\footnote{\url{https://sourceforge.net/projects/heimdall-astro/}}. The data were dedispersed for 295 trial DMs from 0 to 1000~pc~cm$^{-3}$. The trial DMs were chosen such that the S/N of a pulse of 40~ms would result in a S/N drop of 25\% at the next DM trial. Then, the dedispersed time-series were convolved with nine box car functions of widths 256~$\mu$s to 32.768~ms, with increments of powers of two number of samples, and the resulting timeseries was searched for bright pulses with S/N > 6. The candidates above our S/N threshold were then clustered together in DM/arrival time/width parameter space using a brute force clustering algorithm~\citep[see][for more details]{barsdell2012}. The clustering results in a significant reduction in the number of candidates that need to be inspected. The resulting candidates above a S/N of 8 and with more than 5 members in the cluster were then chosen for further inspection.

After the initial filtering of candidates, the final candidate list was passed on to a convolutional neural network candidate classifier called \textsc{FETCH}~\citep{agar2019}. The algorithm uses the time-frequency information and DM-time information of each candidate to classify between RFI and real candidates. The classifier is meant to be used directly without any retraining required on the dataset. To make sure of this, we put all our detected pulses of PSR B0329+54 above a S/N of 8 through the classifier and there were no false positives reported by the algorithm. Hence, we decided to use \textsc{FETCH} without any retraining on our dataset. Finally, the candidates that were classified as true astrophysical candidates by the algorithm were selected for visual inspection. All potential candidates from \textsc{FETCH} were vetted and inspected for further signs of astrophysical origin i.e the broadband pulsed emission and a bow-tie in the DM-time plot.

\begin{figure*}
        \centering
        \subfloat{\includegraphics[width=8cm, height=12cm]{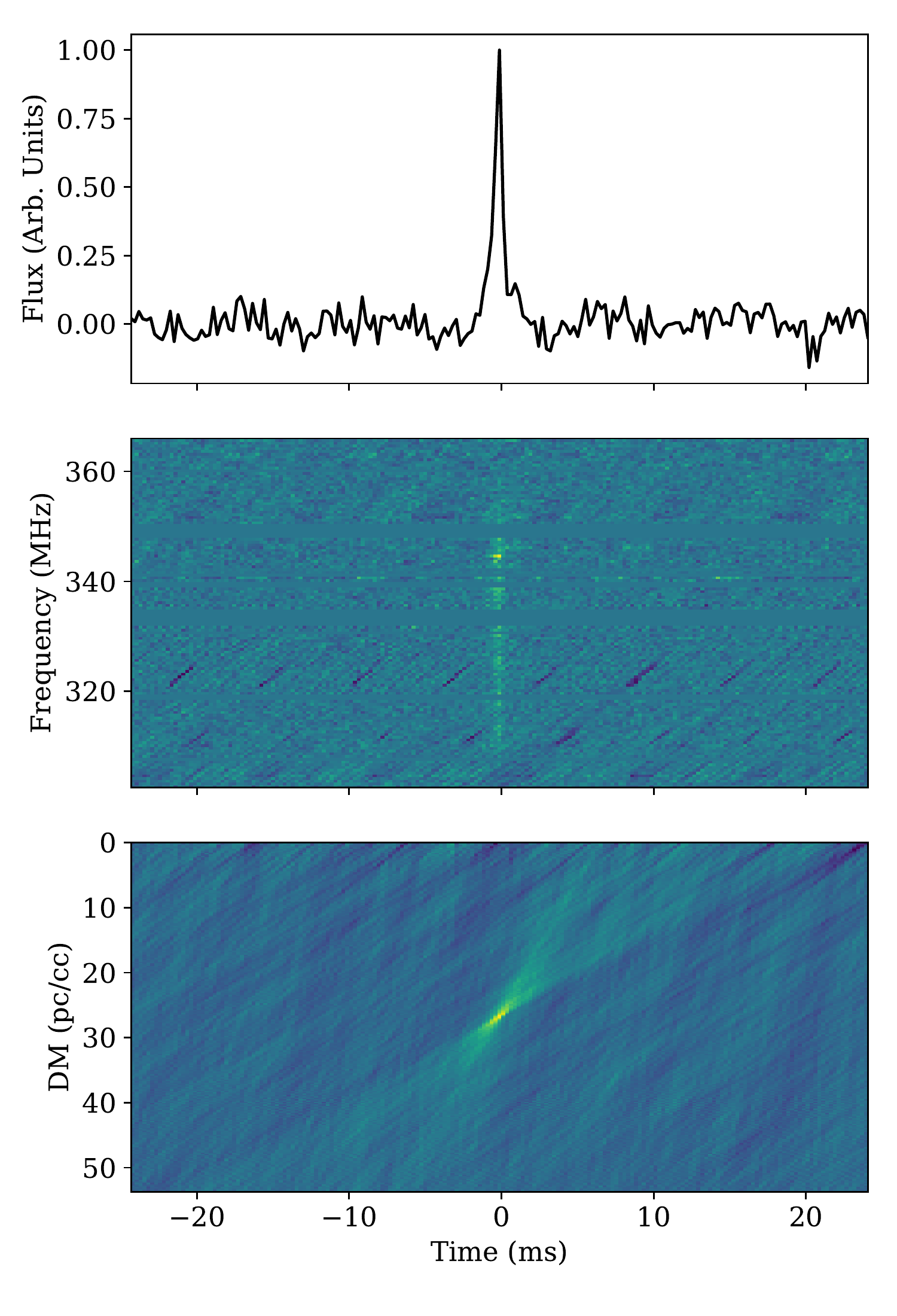}} \subfloat{\includegraphics[width=8cm, height=12cm]{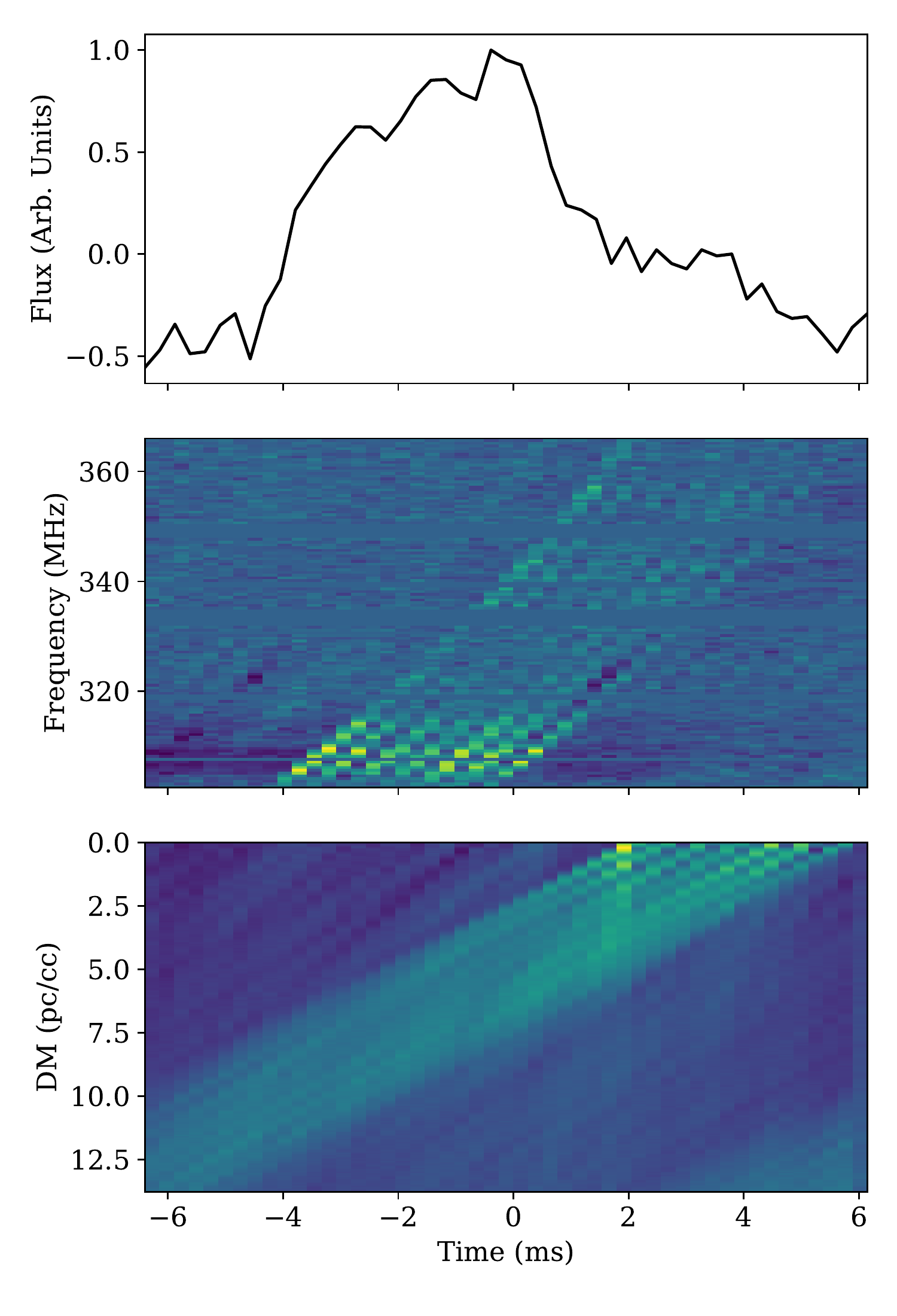}}
\caption{\textbf{Left Column}: Detection of a 23$\sigma$ pulse of PSR B0329+54 on MJD 57914 that was identified by FETCH as a potential candidate. The top panel shows the timeseries of the pulse that has been decimated by a factor of 16 and dedispersed to a DM of 26.7~pc~cm$^{-3}$. Middle panel shows the dedispersed dynamic spectrum of the pulse. Blank channels denote the frequency channels flagged because of RFI. Bottom panel shows the DM-time plot for the same pulse decimated by a factor of 16. \textbf{Right Column}: Similar plots for broadband RFI on the same MJD.}
\label{fig:0329}
\end{figure*}

\section{Results}
\label{section:results}

 Since part of the data were corrupted by RFI, many of the candidates were spurious and terrestrial in origin. Due to the RFI environment, we produced approximately 10$^{3}$ single pulse candidates in every 5 minutes of observations. Using clustering and grouping algorithms in \textsc{heimdall}, we obtained 743915 candidates that were put through \textsc{FETCH}. After further filtering by FETCH, a total of 675 candidates were selected for visual inspection. As mentioned before, PSR B0329+54 drifts through the entire 332~MHz receiver beam hence, we were able to identify 168 single pulses from the source over the span of our search. To make sure our search pipeline was not missing any detectable single pulses, we computed the expected number of single pulses from PSR B0329+54 above a S/N of 8 per traverse. From Figure~\ref{fig:drift}, PSR B0329+54 is detectable above a S/N of 8 in the primary beam for approximately 4 minutes with a mean S/N of 13. Hence, the average single pulse S/N over a span of 4 minutes,
 \begin{equation}
     \rm S/N_{single} = \frac{S/N}{\sqrt{N}}\, \simeq 1,
 \end{equation}
where N is the total number of single pulses. If we assume a log normal distribution of single pulse energies~\citep{burke2012}, we expect about 1 detection per traverse in 4 minutes of observations. Though this is small number, we know that the pulsar is also detectable in the sidelobes and has a large modulation index due to diffractive scintillation at these frequencies~\citep{kr03}. Thus, we expect about 2-3 single pulses from B0329+54 in one complete traverse through the beam. Hence, in 58 days, we expect about 116-180 single pulses which is consistent with the number of pulses we detected.
 Left panel of Figure~\ref{fig:0329} shows one such detection. The rest of the candidates were attributed to instances of strong broadband RFI as shown in the right panel of Figure~\ref{fig:0329}.

Since the survey covered, at any one instance of time, 0.61 sq.deg of sky as calculated from the primary beam and was carried out over 58 days, the non-detection helps us in constraining the FRB rate at 332~MHz. A very simple estimate of the rate can be made by taking the reciprocal of the product of the sky surveyed and the total duration of the survey. Doing that gives us an upper limit of 1166 FRBs per day per sky at 332~MHz. However, we can obtain a more reliable constraint on the upper limit of the rate by considering various factors that affect the true rate of FRBs in any given survey.

\begin{figure}
\centering
\includegraphics[width=\columnwidth]{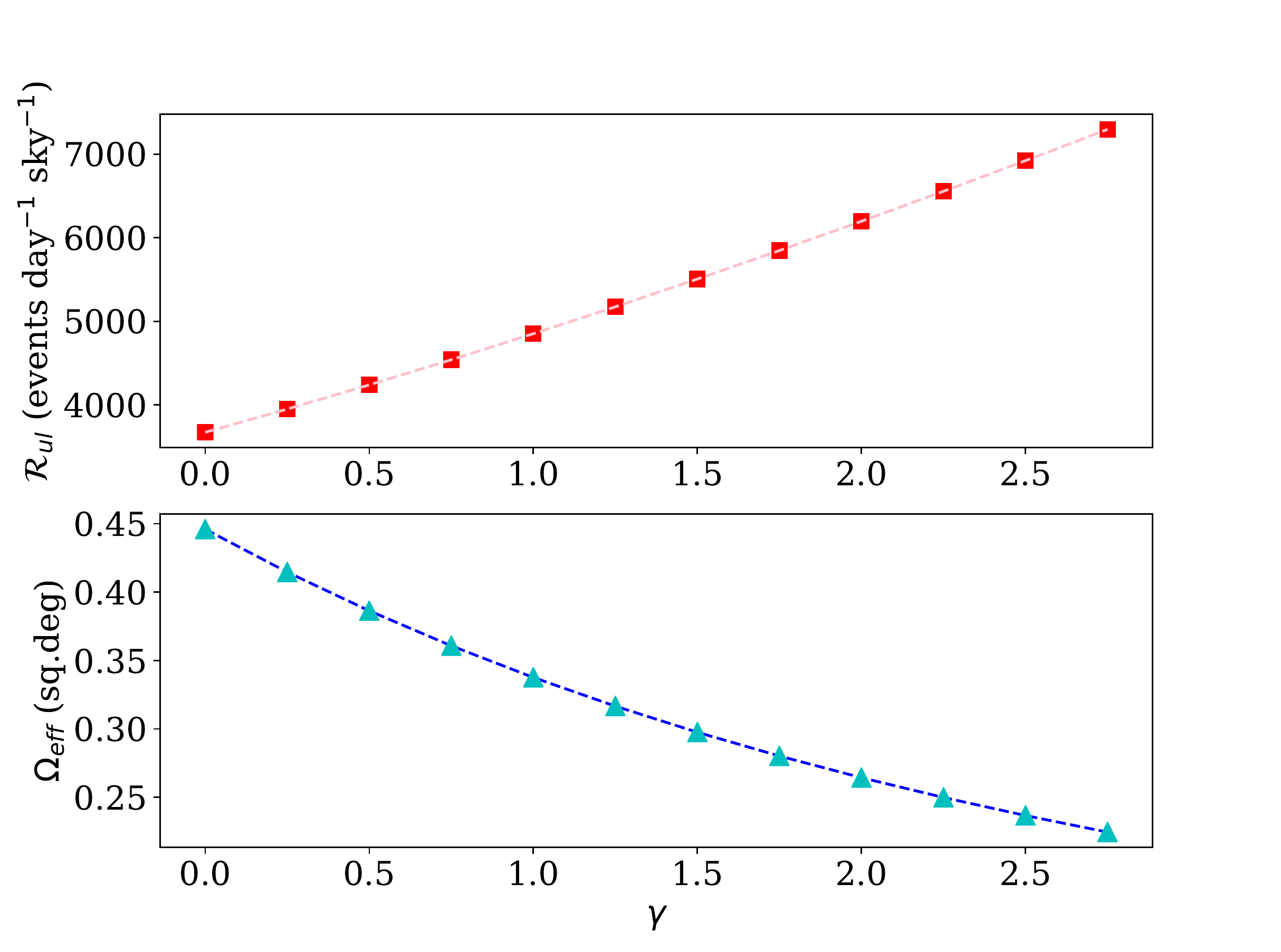}
\caption{The 90$\%$ confidence level upper limit on the rate of FRBs (top panel) and effective FoV (bottom panel) as a function of the slope of the source count distribution. We can see that the constraint on the rate becomes tighter for flatter slopes (see text for more details).}
\label{fig:rate}
\end{figure}

\cite{james2019} have shown that it is important to consider the dependence of the sensitivity of the survey on the offset from the boresight of the beam. Depending on the slope of the source count distribution, the effective FoV of the survey will change for some effective flux threshold. For an effective exposure time $T_{\rm eff}$ and slope of the source count distribution, $\gamma$, the total number of candidates is given by,
\begin{equation}
    N(\gamma) = T_{\rm eff}~\int R(F_{\rm th}, \Omega, \gamma)~B(\Omega) d\Omega,
    \label{eq:N}
\end{equation}
where $B(\Omega)$ is the position dependent beam response and $R(F_{\rm th}, \Omega, \gamma)$ is the event rate above a fluence threshold, $F_{\rm th}$ at a certain offset from the boresite of the beam~\citep[see Eq.19 of][]{james2019}. From~\cite{ch17}, the rate at the boresite of the beam,
\begin{equation}
    R_{\rm 0} = R_{\rm th} \left(\frac{F_{\rm 0}}{F_{\rm th}}\right)^{\gamma},
    \label{eq:rate}
\end{equation}
for a given fluence threshold at boresight, $F_{\rm0}$. Thus, the threshold at any position offset from the boresight,
\begin{equation}
    F_{\rm th}(\Omega) = \frac{F_{\rm 0}}{B(\Omega)}.
    \label{eq:fth}
\end{equation}
Using Eq.~\ref{eq:N}, Eq.~\ref{eq:fth}, and Eq.~\ref{eq:rate}, we can further solve for the rate at the boresite of the beam, $R_{\rm 0}$ to get,
\begin{equation}
    R_{0} = \frac{N(\gamma)}{T_{\rm eff}~\Omega_{\rm eff}} \, ,
\end{equation}
where,
\begin{equation}
    \Omega_{\rm eff} (\gamma) = \int B(\Omega)^{1 - \gamma} d\Omega .
\end{equation}
Then, we can obtain the 90$\%$ confidence level upper limit on the rate for a given $\gamma$ using techniques in~\cite{geh86}. For a given effective FoV, the 90$\%$ upper limit on the rate,
\begin{equation}
\mathcal{R}_{ul} = \frac{2.303}{\Omega_{\rm eff}(\gamma)~T_{\rm eff}}.
\end{equation}
Figure~\ref{fig:rate} shows our limits on the rate as a function of $\gamma$. We should note that we only consider the FoV up to the FWHM of the primary beam in our analysis. Figure~\ref{fig:beam} clearly shows that we are much more sensitive in the sidelobes compared to the ideal beam shape and this will significantly affect our rate estimates for lower values of $\gamma$~\citep[see][for more details]{james2019}. For a uniform distribution of FRBs in a Euclidean Universe ($\gamma=1.5$), our 90$\%$ confidence limit on the rate is $\sim$5500 FRBs per day per sky.

\section{Low Frequency Suppression}
\label{sec:MC}
\subsection{Suppression of Radio Emission}

\begin{figure*}
        \centering
        \subfloat[$\gamma$=1.0]{\includegraphics[scale=0.34]{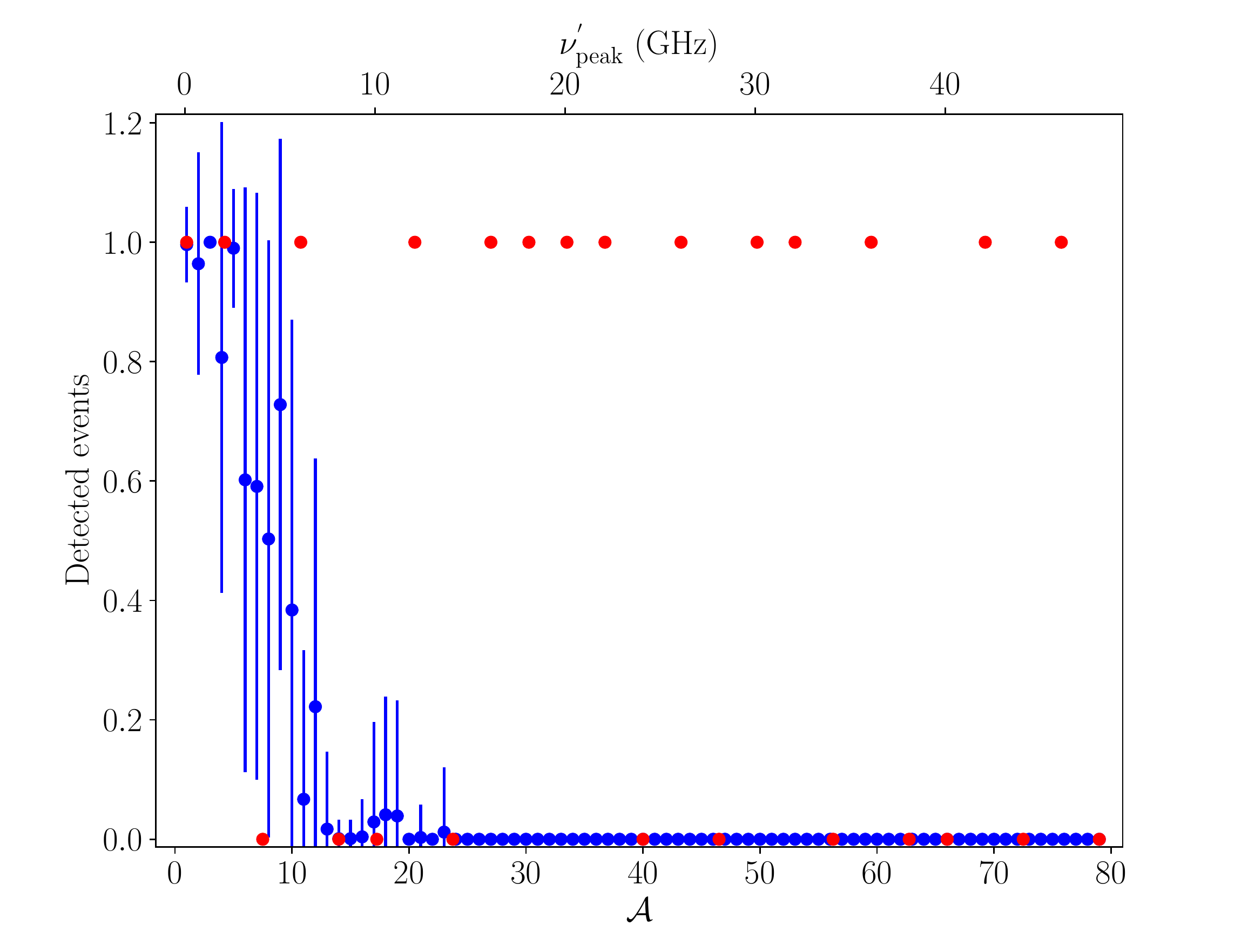}}
        \subfloat[$\gamma$=1.5]{\includegraphics[scale=0.34]{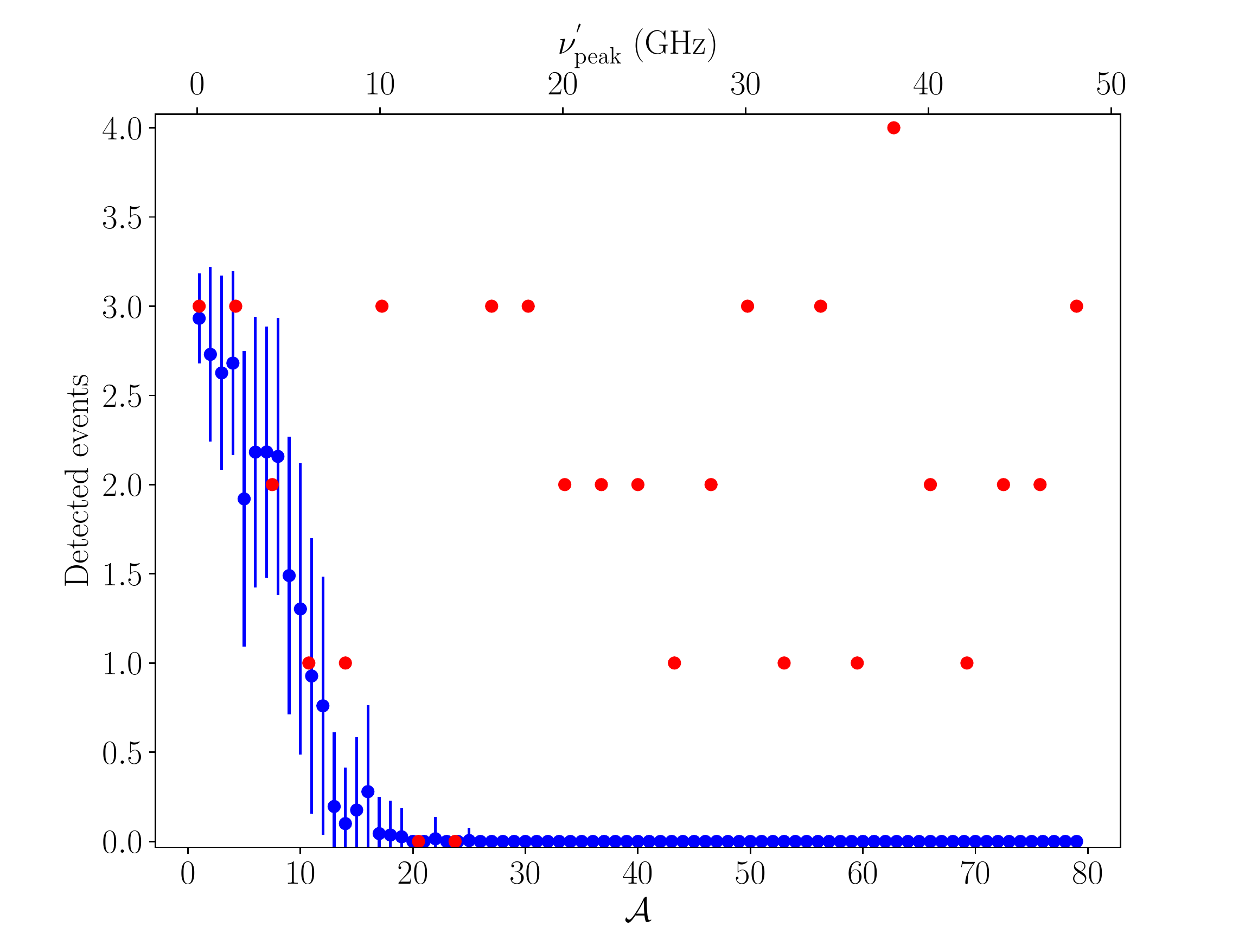}}
        \\
        \subfloat[$\gamma$=2.0]{\includegraphics[scale=0.34]{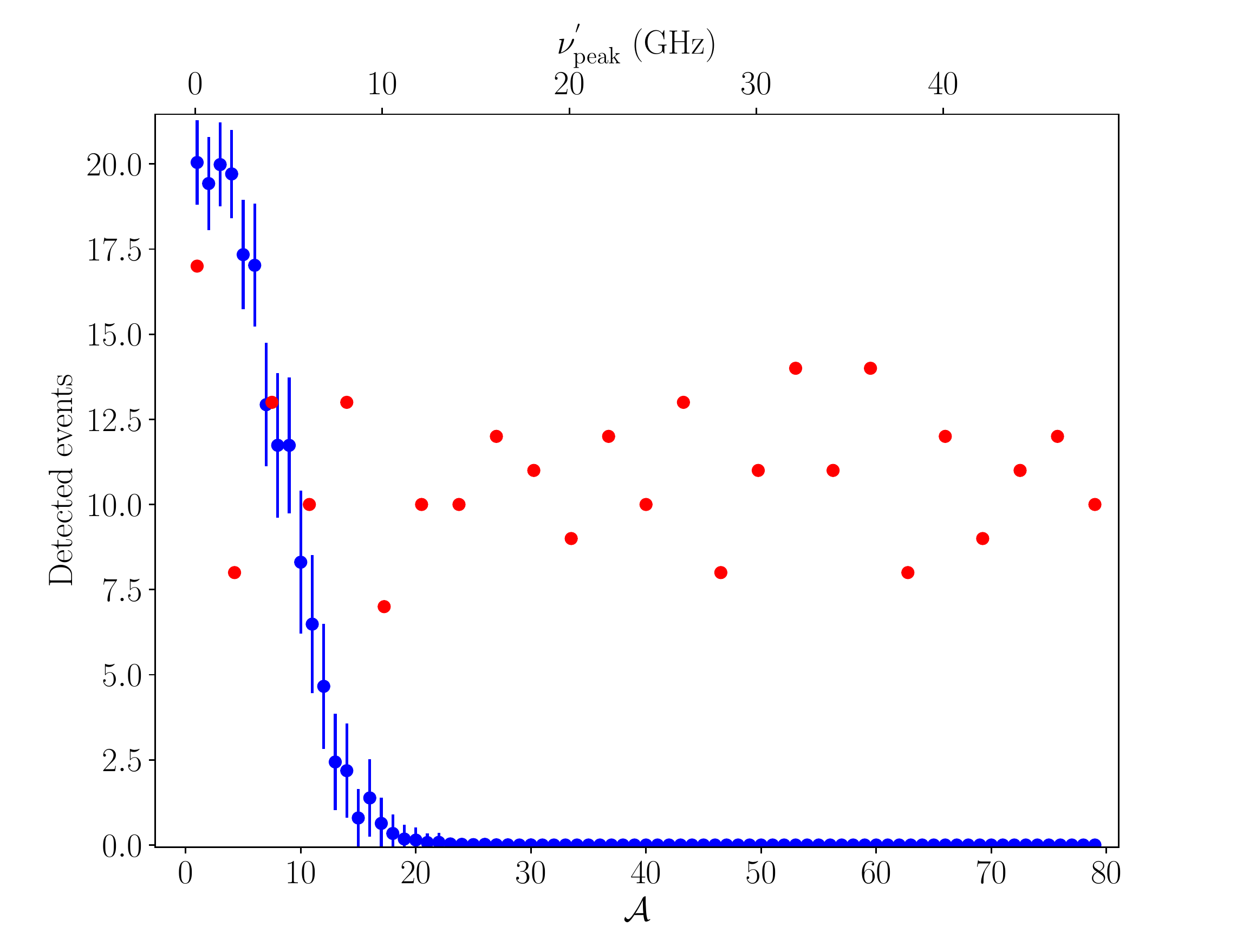}}
        \subfloat[$\gamma$=2.5]{\includegraphics[scale=0.34]{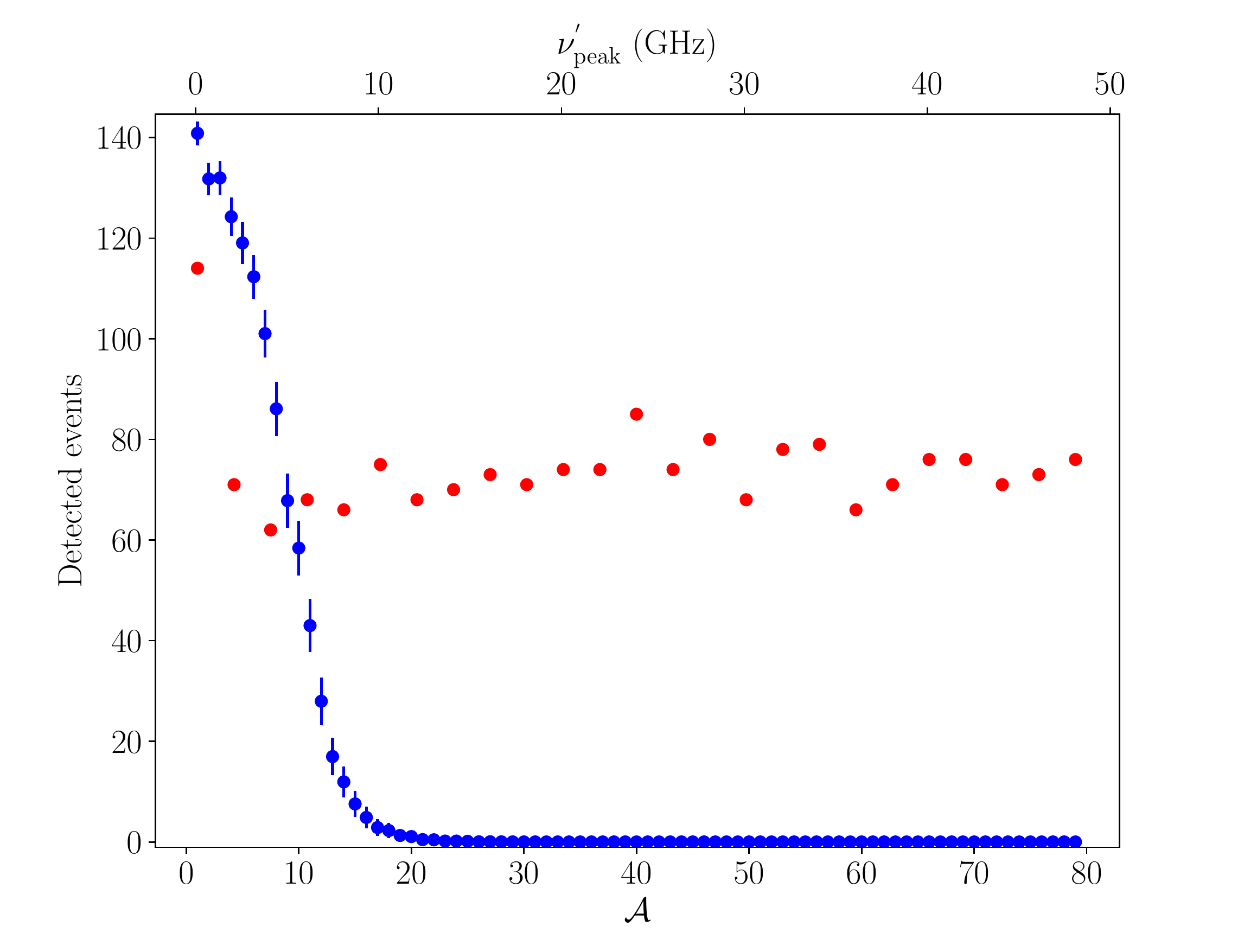}}
        \caption{Number of detections as a function of model parameters. Red points denote the ICS case while blue points denote the FF case. Different plots are for different values of $\gamma$. The discrete values of red points in the simulation are a direct manifestation of the analytical implementation of the ICS model (see text for details). The scatter in red points can be explained by random sampling noise in the MC simulation due to small number of expected detections.} 
        \label{fig:simres}
    \end{figure*}


As mentioned before, there have been quite a few low-frequency searches for FRBs and apart from CHIME~\citep{CHIME2018}, none of them have so far detected FRBs at low frequencies. Non-detection in Lovell, GBNCC and LOFAR surveys combined with a significant number of discoveries in the CHIME band could mean that the peak in the  observed FRB count rate as a function of observing frequency could be due to suppression of FRB emission by external agents. So far, the rate calculations for FRBs assume a power-law distribution in the intrinsic fluence distribution of FRBs but if absorption plays an important role at low frequencies, the estimated upper limit on the rate will be overestimated~\citep[see][for a detailed treatment]{ravi2018a}. Various mechanisms can play a significant role in the vicinity of FRB progenitors and cause severe chromatic absorption and scattering of radio waves, thus altering the intrinsic spectrum of the source. If one assumes that the non-detections at low frequencies are due to absorption, one can obtain constraints on the physical characteristics of the absorbing medium. Thus, one can examine the feasibility of different absorption models based on the estimated constraints and probe the true cause of absorption at these frequencies. In order to achieve this, we used two models discussed in~\cite{ravi2018a}, namely Induced Compton Scattering (henceforth, ICS) and Free-Free absorption (henceforth, FF).

ICS has been used as a possible explanation for low frequency turnovers observed in quasars~\citep{sun1971}. 
If ICS is the dominant absorption process near the source, one observes a bend in the flux density spectrum of the source at a frequency $\nu'_{\rm peak}$ in the rest frame of the source. This manifests itself as a change in the observed spectral index, $\alpha$, such that
\[
    \alpha = 
    \begin{dcases} 
      \alpha & \nu' > \nu'_{\rm peak} \\
      1 -\frac{\alpha}{2} &\nu' < \nu'_{\rm peak} .
    \end{dcases}
\]
\noindent This $\nu'_{\rm peak}$ can be related to the electron density of the absorbing medium. In the optically thin regime, for an intrinsic spectral index $\alpha$, the electron density,
\begin{equation}
n_{\rm e} < \frac{1}{1+\alpha} \frac{m_{\rm e} c^{2}}{k_{b}T_{\rm b}(\nu'_{\rm peak})\sigma_{\rm T}R},
\label{eq:ics}
\end{equation}
where, $\sigma_{\rm T}$ is the Thompson scattering cross-section, $R$ is the size of the scattering medium, $m_{\rm e}$ is the mass of the electron, $k_{b}$ is the Boltzmann constant, $T_{\rm b}$ is the brightness temperature of the source at the peak frequency and $c$ is the speed of light. Thus, constraints on the peak frequency of the turnover can give us a constraint on the electron density around the source. 

For the FF scenario, we can model the optical depth of the absorbing medium,
\begin{equation}
\tau_{\rm ff} = 0.082a~\left(\frac{\nu'}{\rm GHz}\right)^{-2.1}~\left(\frac{\rm EM}{\rm cm^{-6}~pc}\right)\left(\frac{T_{\rm e}}{\rm K}\right)^{-1.35},
\label{eq:ff}
\end{equation}
where the emission measure,
\begin{equation}
{\rm EM} = \int_0^l \langle n_{e}^{2}\rangle~dl,
\end{equation}
where $l$ is the size of the absorbing medium along the line of sight and $a$ is a parameter of the order of unity. Here, $T_{\rm e}$ is the electron temperature~\citep[see][for more details]{ra17} and $\nu'$ is the frequency in the rest frame of the source. In this case, we assume that the optical depth towards the source is dominated by the circum-burst medium with negligible contribution from the Inter-Galactic Medium (IGM) ($\tau_{\rm IGM} \lll$ 1) and hence the measured EM would only come from the absorber. Using reasonable estimates of the electron temperature and linear size of the absorbing medium based on known absorbers in the Galaxy, we can constrain the electron density of the absorber. 

\subsection{Monte-Carlo Simulations}
To obtain meaningful constraints, we first had to obtain a reliable estimate of the `true' event rate at 332~MHz. One way to compute this rate would be to scale the GBNCC rate to our sensitivity~\citep[Eq.11 from][]{ch17}. The caveat here is that the rate scaled from GBNCC could be biased because of flux mitigation due to absorption. Hence, we used the reported rate at 1.4~GHz of 37 FRBs per day per sky from~\cite{sh2018} as our reference rate. The underlying assumption here is that since ASKAP is detecting FRBs at a much higher frequency, the effects of absorption and scattering are minimal and the rate they calculate based on their detections is closer to the `true' rate of events in their frequency range. To account for scattering, we assumed a scattering timescale of $\sim$5~ms based on the average value reported in~\cite{CHIME2018} scaled to 332~MHz using the $\nu^{-4}$ scaling relation with frequency~\citep{cor16}. We ran 1000 realizations of an MC simulation where we picked N events from a uniform distribution of redshifts between 0.05 and 3 and drew their energies from a power-law FRB energy distribution at 1.4~GHz for some $\gamma$ where N depended on the number of events expected from the rate scaled from the ASKAP rate for the given $\gamma$. Here, the FRB energy distribution means the distribution of energies of one-off FRBs and not the distribution of energies of pulses from the same source. This distribution is analogous to the source count distribution~\citep{mac2018}. We note that we used our reported sensitivity flux threshold in the scaling as that estimate accounts for the reduction in sensitivity of the survey due to RFI. We assume that the simulated FRBs emit over a broad range of frequencies. Since we only searched for events up to a DM of 1000~pc~cm$^{-3}$, from the N detected events, we only chose those detected events that were below a redshift of 1 for the rest of the simulation. The redshift limit of 1 was chosen based on the assumption that we spend more than 80$\%$ of the time out of the  Galactic plane with the median DM contribution by the Milky Way of 69~pc~cm$^{-3}$ for all the pointings (see bottom panel of Figure~\ref{fig:dm} for reference). The limits of the energy distribution were adopted from~\cite{ca2019}. Then, we converted the energy of each event to a flux based on the redshift of the FRB. We scaled the obtained 1.4-GHz flux density to 332~MHz and applied reduction to the peak flux due to scattering and the correction due to free-free absorption and ICS in the rest frame of the source for a set of model parameters. For the flux scaling, we used a spectral index that was drawn from a normal distribution with a mean of $-$1.8 and unity variance~\citep{jank2018}. Finally, we obtained the final peak fluxes at 332~MHz for our N simulated events in the observer's frame and ran a simulated search, accounting for the reduction in sensitivity due to channel flagging and RFI over the resulting events. For the ICS case, we ran the simulations for different values of $\nu'_{\rm peak}$ ranging from 0.1 to 50~GHz. For the FF case, we used $\mathcal{A} = {\rm EM}~T_{\rm e}^{-1.35}$ as the model parameter and ran the simulations for $\mathcal{A}$ ranging from 1 to 80 based on typical values of EM and T$_{\rm e}$ in absorbing medium from the literature~\citep{ra16,koo2007,Church2002}. 

\begin{figure}
\centering
\includegraphics[width=\columnwidth]{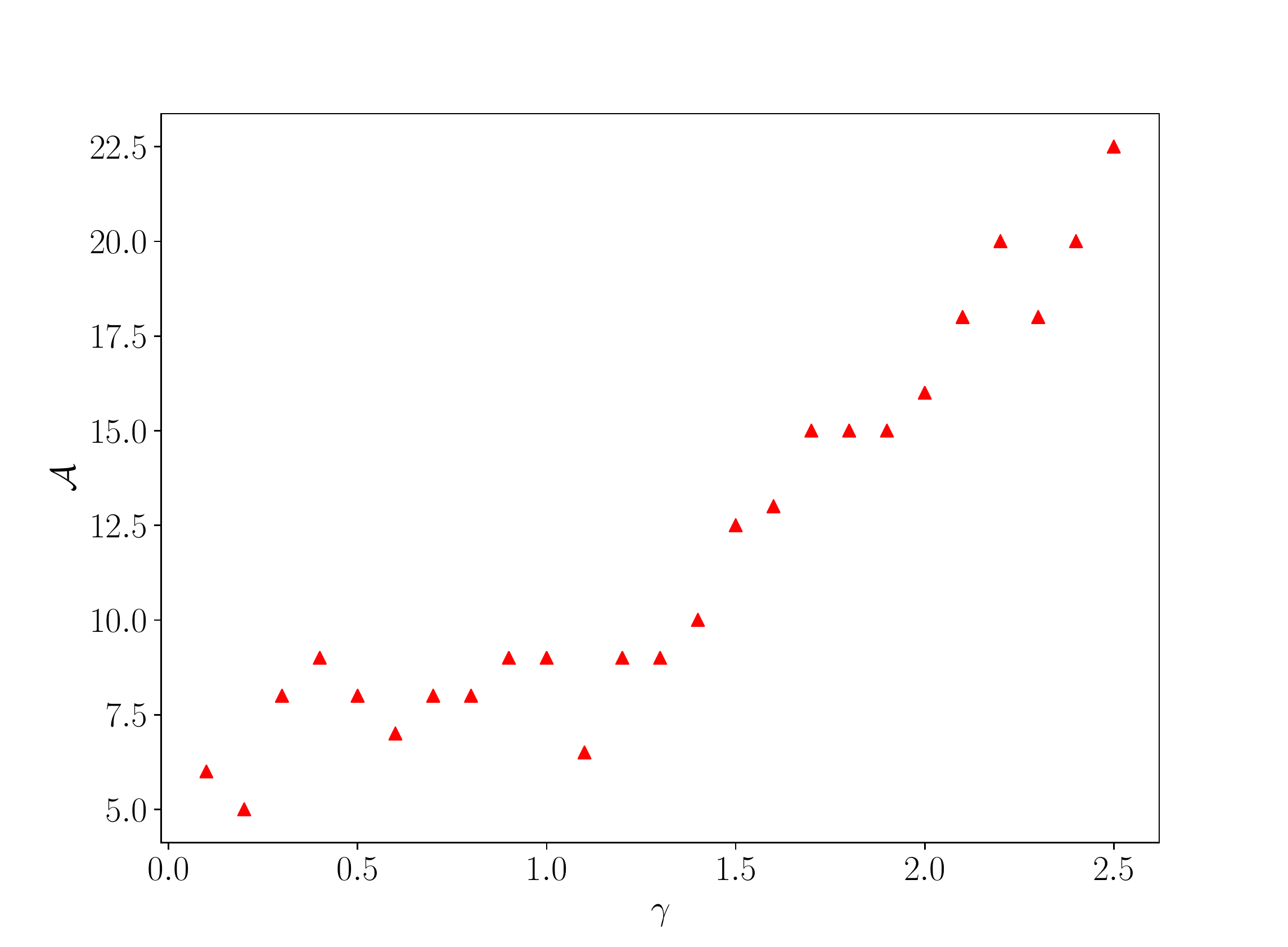}
\caption{Lower limit on the FF model parameter $\mathcal{A}$ as a function of $\gamma$. Note that these are the 95$\%$ upper bounds on the running median values corresponding to zero detections for every $\gamma$. The scatter at lower values of $\gamma$ is due to low number of detections in the simulations.}
\label{fig:gammaA}
\end{figure}

\begin{table}
	\centering
	\caption{Lower limits on the physical parameters of the absorber in our simulations for a range of. $\gamma$s. The limits are defined as values that correspond to zero FRB detections in our simulated survey. The limits were derived from the 95$\%$ upper bound on the measured running median of the 5 successive values of the parameter (see text for more details). The results for ICS case are not shown as our simulations never converged for any value of $\gamma$ (see text for detail).}
	\begin{tabular}{lr} 
		\hline
        Free-Free Absorption (FF) \\
        \hline
        $\gamma$ & $\mathcal{A}$ \\
        0.5 &  8.0\\ 
        1.0 &  9.0\\
        1.5 &  12.0\\
        2.0 &  17.0\\
        2.5 &  22.0\\
        \hline
	\end{tabular}
	\label{tab:ll}
\end{table}
We ran the MC simulations described above for $\gamma$s in the range of 0.1--2.5. Figure~\ref{fig:simres} shows the number of detections as a function of model parameters for four different values of $\gamma$. Since we were working with low number of detections, we expected to hit random noise variations due to the sampling from the MC simulations resulting in some erratic variations in number of detections with decreasing $\gamma$ (see Figure.~\ref{fig:simres}). Here, we report non-integer number of detections as the fractional part reflects the probability of finding another event for the given event rate, sky coverage and time on sky. In order to obtain the proper lower limits on the parameters, we used a running median of number of detected events across ten successive values of  $\mathcal{A}$ for the FF case and $\nu'_{\rm peak}$ for the ICS case. Then the lower limit on $\mathcal{A}$ or $\nu'_{\rm peak}$ is the first value for which the running median of the number of detections is less than 0.5. Finally, we use the 95$\%$ confidence level upper bound on this measured value of the running median to calculate the lower limit on $\mathcal{A}$. The corresponding values for the lower limit on $\mathcal{A}$ for different values of $\gamma$s are shown in Table~\ref{tab:ll}. For the FF case, we observe that the lower limit on $\mathcal{A}$ increases with increasing $\gamma$ (see Figure.~\ref{fig:gammaA}) a higher $\gamma$ results in a larger number of expected detections at 332~MHz that in turn means that the optical depth of the absorber has to be higher to render all those events undetectable. For the ICS case, we observe that none of the $\gamma$ values in our selected range gave us zero detections over the entire range of peak frequencies we considered. We discuss the implications of our findings in the next section. 

\section{Discussions and Summary}
\label{sec:discussions}

\begin{table*}
    \centering
	\begin{tabular}{l|l|l|l|l|}
	     $\gamma$ & T$_{e}$ & l & $n_{e}$ & Reference\\
	     & (K) & (pc) & cm$^{-3}$ &  \\
		\hline
        UC HII regions & & & & \\
        \hline
        0.5 & 10000 & 0.03  & 8.1 $\times$ 10$^{3}$ & 1  \\
        1.0 & 10000 & 0.03  & 8.7 $\times$ 10$^{3}$ & 1 \\
        1.5 & 10000 & 0.03  & 1.2 $\times$ 10$^{4}$ &  1 \\
        2.0 & 10000 & 0.03  & 1.3 $\times$ 10$^{4}$ & 1 \\
        2.5 & 10000 & 0.03  & 1.3 $\times$ 10$^{4}$ & 1 \\
        
        \hline
        Ionized SNR filaments  & & & & \\
        \hline
        0.5 & 5000 & 0.015  & 7.2 $\times$ 10$^{3}$ & 2\\
        1.0 & 5000 & 0.015  & 7.6 $\times$ 10$^{3}$ & 2\\
        1.5 & 5000 & 0.015  & 8.8 $\times$ 10$^{3}$ & 2\\
        2.0 & 5000 & 0.015  & 1.0 $\times$ 10$^{4}$ & 2\\
        2.5 & 5000 & 0.015  & 1.2 $\times$ 10$^{4}$ & 2\\
        \hline
         Post-shock region (SLSNe)   & & & & \\
        \hline
        0.5 & 10$^{7}$ & 0.03 & 2.7 $\times$ 10$^{5}$ & 3\\
        1.0 & 10$^{7}$& 0.03  & 2.9 $\times$ 10$^{5}$ & 3\\
        1.5 & 10$^{7}$ & 0.03 & 1.0 $\times$ 10$^{6}$ & 3\\
        2.0 & 10$^{7}$ & 0.03 & 1.2 $\times$ 10$^{6}$ & 3\\
        2.5 & 10$^{7}$& 0.03  & 1.4 $\times$ 10$^{6}$ & 3\\
        \hline
	\end{tabular}
	\caption{Values of physical characteristics of the absorbing medium. For the UC HII and
	Ionized SNR scenario, we use the upper limit on the linear size as reported
	in~\citet{sok2018} and for the SLSNe, we use the upper limit on the linear size of the absorber from~\citet{mar2018}. The nominal values for T$_{e}$ and EM are taken from \citet{Church2002} (1), \citet{koo2007} (2) and \citet{mar2018} (3). The values for $n_{e}$ are reported from the analysis in this paper.}
	\label{tab:par}
\end{table*}

\begin{figure}
\includegraphics[scale=0.44]{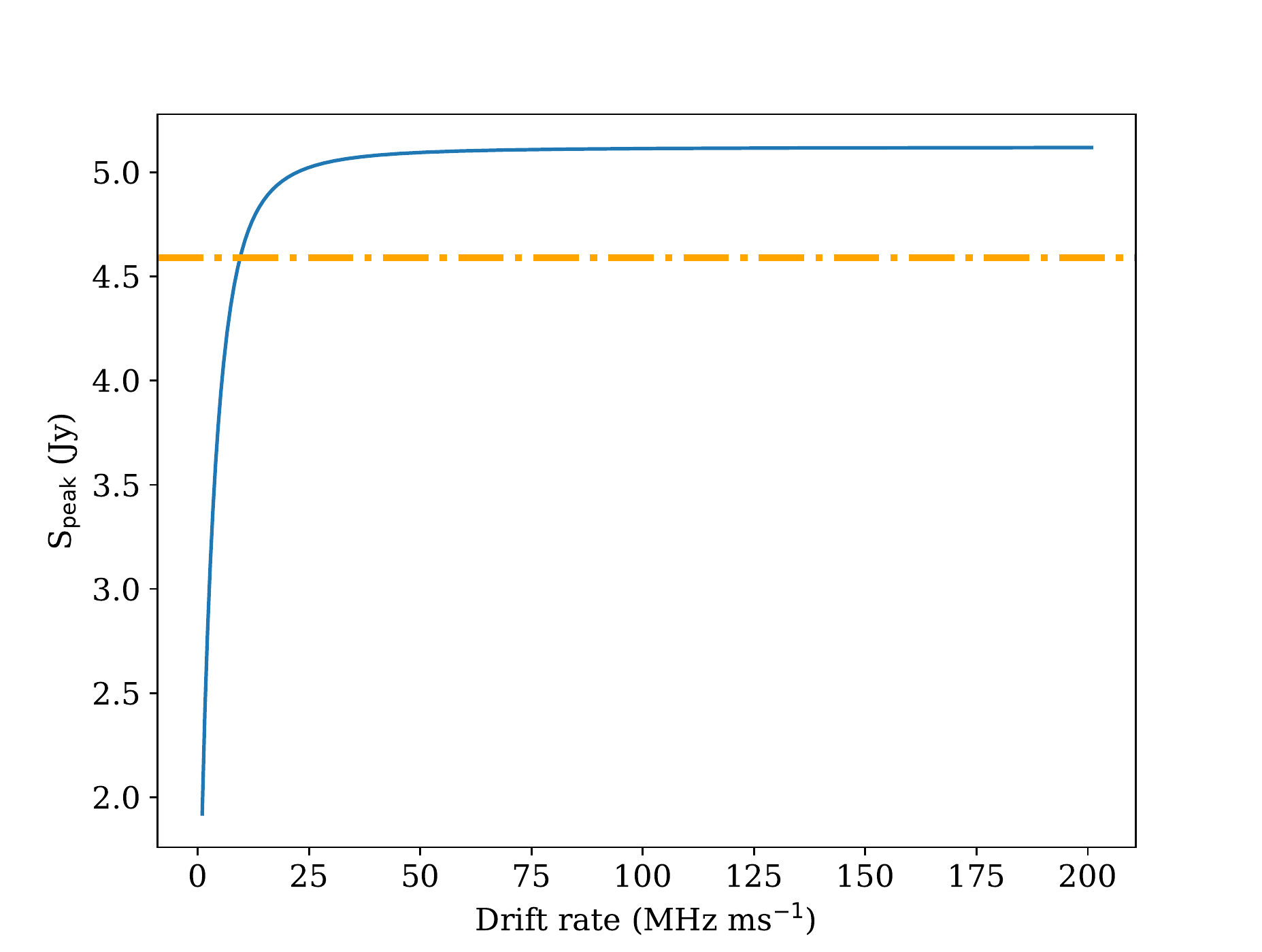}
\caption{Peak flux of a simulated FRB at 332 MHz as a function of drift rate. The orange line corresponds to the peak flux sensitivity of our survey.}
\label{fig:drift}
\end{figure}

The dearth at FRBs detected at the lowest radio frequencies might suggest that there is either something intrinsic to the emission mechanism or it could be due to propagation effects. The survey presented here allows us to consider the latter. The existence of frequency drifting in FRB 121102 and a few of the other repeating sources discovered by CHIME~\citep{CHIME2019c} poses an interesting question of whether the intrinsic spectral characteristics of these FRBs could be responsible for rendering them undetectable at lower frequencies.~\cite{jos2019} detected a pulse from FRB 121102 at 600~MHz and calculated a drift rate of 3.9 MHz per ms compared to 200 MHz per ms at 1.4~GHz reported in~\cite{hes2019}. This means that pulses at even lower frequencies might get completely smeared out by the frequency drifting phenomenon as shown in Figure.~\ref{fig:drift}. Though the question of whether frequency drifting sub-pulses is a property inherent to all FRBs still remains to be answered, it will definitely affect the peak flux of a fraction of the sources.

The lack of detections in previous low frequency surveys combined with detections of FRBs in the 400--800~MHz band might suggest that absorption of radio emission at low frequency could explain the observations.~\cite{sok2018} were able to provide sensible limits on the absorbing medium based on the non-detection of 7 FRBs that were detected by ASKAP simultaneously. They report an upper limit on the size of the absorbing medium to be 0.03~(T$_{e}$/10$^{4}$~K)~pc. If we consider various examples of absorbers that are found in our Galaxy, we can use the reported upper limit on the linear size to constrain the most plausible circum-burst medium that would be applicable to FRBs. To do that, we considered three cases; 1) Ultra-Compact (UC) HII regions where electron densities can be of the order of 10$^{4}$ cm$^{-3}$~\citep{Church2002}; 2) dense ionized filaments that are found in Supernova Remnants (SNR)~\citep{koo2007}; and 3) Post-shock region of Super-Luminous Supernovae (SLSNe) where the electron temperature can reach 10$^{6-7}$~K with electron densities of 10$^{5-6}$~cm$^{-3}$~\citep{mar2018}. Using the expected values of electron temperature in these regions, we constrain the electron density of the absorber for lower limits of $\mathcal{A}$ reported in Table~\ref{tab:ll} for different $\gamma$s. For a given T$_{e}$, the lower limit on the electron density,
\begin{equation}
n_{e} = \sqrt{\frac{\mathcal{A}}{{\rm T}_{e}^{-1.35}~l}}\, ,
\end{equation}
where $l$ is the linear size of the absorber along the line of sight. Here, we neglect the fluctuations of electron density within the absorber to get the above relation.
Our results are shown in Table~\ref{tab:par}. While the tabulated electron densities for UC HII regions are not unphysical, based on the electron density-linear size relationship of UC HII regions~\citep{Church2002}, the expected size of the region to sustain such high electron density is a factor of 100 greater than upper limit we used from~\cite{sok2018}. Hence, UC HII regions are unlikely to be the primary source of absorption. The obtained values of electron density for the absorber are larger than the densest filaments that have been observed in Type-II SNRs like   SNR G11.2$-$0.3~\citep{koo2007} and the Crab Nebula~\citep{san98}. ~\cite{mar2018} have proposed a model for FRBs where the pulse originates from a millisecond magnetar that has formed in a SLSN. Immediately following the supernova explosion, extremely high densities and temperatures are reached in the post-shock region of the outflowing ejecta that can cause significant absorption of radio emission from the magnetar over a timescale of a few hundred years at which point, the ejecta have expanded enough to become transparent to low-frequency emission ($\tau \ll$ 1). Based on our MC analysis, the expected electron densities are consistent with what is postulated for these types of environments and can then explain the absorption of radio emission from the FRB progenitor at low frequencies.  

For the ICS scenario, we find that for none of our selected $\gamma$ do we get a non-detection.~\cite{ravi2018a} have emphasized the fact that for all reasonable values of electron density and temperature of the circum-burst plasma, one would expect ICS to be a contributing factor while some special conditions are needed for free-free absorption to be important. Our simulations have shown that for the conditions considered here, ICS alone cannot account for the suppression of radio emission at low frequencies though a combination of ICS and FF processes can result in the observed behaviour. We also cannot rule out a completely different absorbing mechanism to the ones that have been proposed so far~\citep[see][for a further review]{ravi2018a}. The detection of FRBs by CHIME along with our non-detections can place tighter constraints on the absorbing medium. The frequency range of CHIME is ideal to detect a turnover in the FRB event rate if it exists and will be able tell us more about the absorption happening along the these lines of sights. Since, CHIME is still in the process of calibrating their sensitivity and their beam shape, we cannot use the present results to obtain any meaningful constraints from their discoveries.

In summary, we have carried out a survey for FRBs at 332~MHz. We searched a total of 58 days of observations and found no FRBs in our survey. Based on the non-detections, our 90$\%$ confidence upper limit on the rate varies from 4000--7500 events per day per sky above a flux threshold of 4.6~Jy depending on the value of $\gamma$ we use. We discuss various possibilities for non-detections at lower frequencies and show that absorption of radio emission can definitely result in non-detections though intrinsic spectral characteristics and sub-pulse frequency drifting can affect detection rates to some extent. Using our non-detection, we constrain the physical mechanisms of absorption of radio emission. From our MC simulations, we show that Induced Compton Scattering alone cannot account for the entire absorption of emission at these frequencies. We go on to constrain the mean electron density of the medium assuming Free-Free absorption to be the primary source of absorption. We find that the limits are inconsistent with dense ionized filaments that are prevalent in SNRs or with dense UC-HII regions along the line of sight. However, our constraints do agree with out-flowing dense ejecta within a SLSN, a model that was proposed by~\cite{mar2018} as a site for FRB progenitors. This, if indeed true, provides further evidence for a compact object origin of FRBs. More detections of FRBs with CHIME at 600~MHz and the UHF-band at MeerKAT radio telescope will enable better understanding of the low frequency spectrum of these enigmatic sources.

\section*{Acknowledgements}
KMR and BWS acknowledge support from the European Research Council (ERC) under the European Union's Horizon 2020 research and innovation programme (grant agreement No 694745). RPB acknowledges support from the European Research Council under the European Union's Horizon 2020 research and innovation programme (grant agreement no. 715051; Spiders). KMR thanks Rohini Joshi for kindly sharing the GPU-based MAD filter code for our use. KMR would like to thank Devansh Agarwal for providing assistance with running \textsc{FETCH} on our search data.



\bibliographystyle{mnras}
\bibliography{newbib2} 

\begin{thebibliography}{}
\makeatletter
\relax
\def\mn@urlcharsother{\let\do\@makeother \do\$\do\&\do\#\do\^\do\_\do\%\do\~}
\def\mn@doi{\begingroup\mn@urlcharsother \@ifnextchar [ {\mn@doi@}
  {\mn@doi@[]}}
\def\mn@doi@[#1]#2{\def\@tempa{#1}\ifx\@tempa\@empty \href
  {http://dx.doi.org/#2} {doi:#2}\else \href {http://dx.doi.org/#2} {#1}\fi
  \endgroup}
\def\mn@eprint#1#2{\mn@eprint@#1:#2::\@nil}
\def\mn@eprint@arXiv#1{\href {http://arxiv.org/abs/#1} {{\tt arXiv:#1}}}
\def\mn@eprint@dblp#1{\href {http://dblp.uni-trier.de/rec/bibtex/#1.xml}
  {dblp:#1}}
\def\mn@eprint@#1:#2:#3:#4\@nil{\def\@tempa {#1}\def\@tempb {#2}\def\@tempc
  {#3}\ifx \@tempc \@empty \let \@tempc \@tempb \let \@tempb \@tempa \fi \ifx
  \@tempb \@empty \def\@tempb {arXiv}\fi \@ifundefined
  {mn@eprint@\@tempb}{\@tempb:\@tempc}{\expandafter \expandafter \csname
  mn@eprint@\@tempb\endcsname \expandafter{\@tempc}}}

\bibitem[\protect\citeauthoryear{{Agarwal}, {Aggarwal}, {Burke-Spolaor},
  {Lorimer}  \& {Garver-Daniels}}{{Agarwal} et~al.}{2019}]{agar2019}
{Agarwal} D.,  {Aggarwal} K.,  {Burke-Spolaor} S.,  {Lorimer} D.~R.,
  {Garver-Daniels} N.,  2019, arXiv e-prints, \href
  {https://ui.adsabs.harvard.edu/abs/2019arXiv190206343A} {}

\bibitem[\protect\citeauthoryear{{Bannister} et~al.,}{{Bannister}
  et~al.}{2019}]{ban2019}
{Bannister} K.~W.,  et~al., 2019, \mn@doi [Science] {10.1126/science.aaw5903},
  \href {https://ui.adsabs.harvard.edu/abs/2019Sci...365..565B} {365, 565}

\bibitem[\protect\citeauthoryear{{Barsdell}}{{Barsdell}}{2012}]{barsdell2012}
{Barsdell} B.~R.,  2012, PhD thesis, Swinburne University of Technology

\bibitem[\protect\citeauthoryear{{Bassa} et~al.,}{{Bassa}
  et~al.}{2016}]{bassa2016}
{Bassa} C.~G.,  et~al., 2016, \mn@doi [\mnras] {10.1093/mnras/stv2755}, \href
  {https://ui.adsabs.harvard.edu/abs/2016MNRAS.456.2196B} {456, 2196}

\bibitem[\protect\citeauthoryear{{Burke-Spolaor} et~al.,}{{Burke-Spolaor}
  et~al.}{2012}]{burke2012}
{Burke-Spolaor} S.,  et~al., 2012, \mn@doi [\mnras]
  {10.1111/j.1365-2966.2012.20998.x}, \href
  {https://ui.adsabs.harvard.edu/abs/2012MNRAS.423.1351B} {423, 1351}

\bibitem[\protect\citeauthoryear{{CHIME/FRB Collaboration} et~al.,}{{CHIME/FRB
  Collaboration} et~al.}{2018}]{CHIME2018}
{CHIME/FRB Collaboration} et~al., 2018, \mn@doi [\apj]
  {10.3847/1538-4357/aad188}, \href
  {http://adsabs.harvard.edu/abs/2018ApJ...863...48C} {863, 48}

\bibitem[\protect\citeauthoryear{{Caleb} et~al.,}{{Caleb} et~al.}{2017}]{ca17}
{Caleb} M.,  et~al., 2017, preprint

\bibitem[\protect\citeauthoryear{{Caleb}, {Stappers}, {Rajwade}  \&
  {Flynn}}{{Caleb} et~al.}{2019}]{ca2019}
{Caleb} M.,  {Stappers} B.,  {Rajwade} K.,   {Flynn} C.,  2019, arXiv e-prints,
  \href {http://adsabs.harvard.edu/abs/2019arXiv190200272C} {}

\bibitem[\protect\citeauthoryear{{Champion} et~al.,}{{Champion}
  et~al.}{2016}]{ch16}
{Champion} D.~J.,  et~al., 2016, MNRAS

\bibitem[\protect\citeauthoryear{{Chatterjee} et~al.,}{{Chatterjee}
  et~al.}{2017}]{cha17}
{Chatterjee} S.,  et~al., 2017, \nat, 541, 58

\bibitem[\protect\citeauthoryear{{Chawla} et~al.,}{{Chawla}
  et~al.}{2017}]{ch17}
{Chawla} P.,  et~al., 2017, ArXiv e-prints 1701.07457

\bibitem[\protect\citeauthoryear{{Churchwell}}{{Churchwell}}{2002}]{Church2002}
{Churchwell} E.,  2002, \mn@doi [\araa]
  {10.1146/annurev.astro.40.060401.093845}, \href
  {https://ui.adsabs.harvard.edu/abs/2002ARA%26A..40...27C} {40, 27}

\bibitem[\protect\citeauthoryear{{Coenen} et~al.,}{{Coenen}
  et~al.}{2014}]{coe2014}
{Coenen} T.,  et~al., 2014, \mn@doi [\aap] {10.1051/0004-6361/201424495}, \href
  {https://ui.adsabs.harvard.edu/abs/2014A%26A...570A..60C} {570, A60}

\bibitem[\protect\citeauthoryear{{Cordes}, {Wharton}, {Spitler}, {Chatterjee}
  \& {Wasserman}}{{Cordes} et~al.}{2016}]{cor16}
{Cordes} J.~M.,  {Wharton} R.~S.,  {Spitler} L.~G.,  {Chatterjee} S.,
  {Wasserman} I.,  2016, ArXiv e-prints 1605.05890

\bibitem[\protect\citeauthoryear{{Deneva} et~al.,}{{Deneva}
  et~al.}{2016}]{de16}
{Deneva} J.~S.,  et~al., 2016, ApJ, 821, 10

\bibitem[\protect\citeauthoryear{{Gehrels}}{{Gehrels}}{1986}]{geh86}
{Gehrels} N.,  1986, \mn@doi [\apj] {10.1086/164079}, \href
  {https://ui.adsabs.harvard.edu/abs/1986ApJ...303..336G} {303, 336}

\bibitem[\protect\citeauthoryear{Guzm{\'{a}}n, May, Alvarez  \&
  Maeda}{Guzm{\'{a}}n et~al.}{2010}]{Guzmn2010}
Guzm{\'{a}}n A.~E.,  May J.,  Alvarez H.,   Maeda K.,  2010, \mn@doi [Astronomy
  {\&} Astrophysics] {10.1051/0004-6361/200913628}, 525, A138

\bibitem[\protect\citeauthoryear{{Hessels} et~al.,}{{Hessels}
  et~al.}{2019}]{hes2019}
{Hessels} J.~W.~T.,  et~al., 2019, \mn@doi [\apjl] {10.3847/2041-8213/ab13ae},
  \href {https://ui.adsabs.harvard.edu/abs/2019ApJ...876L..23H} {876, L23}

\bibitem[\protect\citeauthoryear{{James} et~al.,}{{James}
  et~al.}{2019}]{james2019}
{James} C.~W.,  et~al., 2019, \mn@doi [\pasa] {10.1017/pasa.2019.1}, \href
  {https://ui.adsabs.harvard.edu/abs/2019PASA...36....9J} {36, e009}

\bibitem[\protect\citeauthoryear{{Jankowski}, {van Straten}, {Keane}, {Bailes},
  {Barr}, {Johnston}  \& {Kerr}}{{Jankowski} et~al.}{2018}]{jank2018}
{Jankowski} F.,  {van Straten} W.,  {Keane} E.~F.,  {Bailes} M.,  {Barr} E.~D.,
   {Johnston} S.,   {Kerr} M.,  2018, \mn@doi [\mnras] {10.1093/mnras/stx2476},
  \href {https://ui.adsabs.harvard.edu/abs/2018MNRAS.473.4436J} {473, 4436}

\bibitem[\protect\citeauthoryear{Jones, Davis, Wilkinson, Giardino, Melhuish,
  Asareh, Davies  \& Lasenby}{Jones et~al.}{2001}]{Jones2001}
Jones A.~W.,  Davis R.~J.,  Wilkinson A.,  Giardino G.,  Melhuish S.~J.,
  Asareh H.,  Davies R.~D.,   Lasenby A.~N.,  2001, \mn@doi [Monthly Notices of
  the Royal Astronomical Society] {10.1046/j.1365-8711.2001.04771.x}, 327, 545

\bibitem[\protect\citeauthoryear{{Josephy} et~al.,}{{Josephy}
  et~al.}{2019}]{jos2019}
{Josephy} A.,  et~al., 2019, \mn@doi [\apjl] {10.3847/2041-8213/ab2c00}, \href
  {https://ui.adsabs.harvard.edu/abs/2019ApJ...882L..18J} {882, L18}

\bibitem[\protect\citeauthoryear{{Karastergiou} et~al.,}{{Karastergiou}
  et~al.}{2015}]{ka15}
{Karastergiou} A.,  et~al., 2015, MNRAS, 452, 1254

\bibitem[\protect\citeauthoryear{{Koo}, {Moon}, {Lee}, {Lee}  \&
  {Matthews}}{{Koo} et~al.}{2007}]{koo2007}
{Koo} B.-C.,  {Moon} D.-S.,  {Lee} H.-G.,  {Lee} J.-J.,   {Matthews} K.,  2007,
  \mn@doi [\apj] {10.1086/510550}, \href
  {https://ui.adsabs.harvard.edu/abs/2007ApJ...657..308K} {657, 308}

\bibitem[\protect\citeauthoryear{{Kramer}, {Karastergiou}, {Gupta}, {Johnston},
  {Bhat}  \& {Lyne}}{{Kramer} et~al.}{2003}]{kr03}
{Kramer} M.,  {Karastergiou} A.,  {Gupta} Y.,  {Johnston} S.,  {Bhat} N.~D.~R.,
    {Lyne} A.~G.,  2003, A\&A, 407, 655

\bibitem[\protect\citeauthoryear{{Kulkarni}, {Ofek}  \& {Neill}}{{Kulkarni}
  et~al.}{2015}]{ku15}
{Kulkarni} S.~R.,  {Ofek} E.~O.,   {Neill} J.~D.,  2015, ArXiv e-prints
  1511.09137

\bibitem[\protect\citeauthoryear{{Lorimer} \& {Kramer}}{{Lorimer} \&
  {Kramer}}{2004}]{lo04}
{Lorimer} D.~R.,  {Kramer} M.,  2004, {Handbook of Pulsar Astronomy}

\bibitem[\protect\citeauthoryear{{Lorimer}, {Bailes}, {McLaughlin}, {Narkevic}
  \& {Crawford}}{{Lorimer} et~al.}{2007}]{lo07}
{Lorimer} D.~R.,  {Bailes} M.,  {McLaughlin} M.~A.,  {Narkevic} D.~J.,
  {Crawford} F.,  2007, Science, 318, 777

\bibitem[\protect\citeauthoryear{{Macquart} \& {Ekers}}{{Macquart} \&
  {Ekers}}{2018}]{mac2018}
{Macquart} J.-P.,  {Ekers} R.~D.,  2018, \mn@doi [\mnras]
  {10.1093/mnras/stx2825}, \href
  {http://adsabs.harvard.edu/abs/2018MNRAS.474.1900M} {474, 1900}

\bibitem[\protect\citeauthoryear{{Margalit} \& {Metzger}}{{Margalit} \&
  {Metzger}}{2018}]{mar2018}
{Margalit} B.,  {Metzger} B.~D.,  2018, \mn@doi [\apjl]
  {10.3847/2041-8213/aaedad}, \href
  {https://ui.adsabs.harvard.edu/abs/2018ApJ...868L...4M} {868, L4}

\bibitem[\protect\citeauthoryear{{Petroff} et~al.,}{{Petroff}
  et~al.}{2015}]{pe15}
{Petroff} E.,  et~al., 2015, MNRAS, 447, 246

\bibitem[\protect\citeauthoryear{{Press}, {Flannery}, {Teukolsky}  \&
  {Vetterling}}{{Press} et~al.}{1989}]{pr89}
{Press} W.~H.,  {Flannery} B.~P.,  {Teukolsky} S.~A.,   {Vetterling} W.~T.,
  1989, {Numerical recipes in C. The art of scientific computing}.
Cambridge: University Press

\bibitem[\protect\citeauthoryear{{Rajwade} \& {Lorimer}}{{Rajwade} \&
  {Lorimer}}{2017}]{ra17}
{Rajwade} K.~M.,  {Lorimer} D.~R.,  2017, \mnras, 465, 2286

\bibitem[\protect\citeauthoryear{{Rajwade}, {Lorimer}  \& {Anderson}}{{Rajwade}
  et~al.}{2016}]{ra16}
{Rajwade} K.,  {Lorimer} D.~R.,   {Anderson} L.~D.,  2016, MNRAS, 455, 493

\bibitem[\protect\citeauthoryear{{Ravi} \& {Loeb}}{{Ravi} \&
  {Loeb}}{2018}]{ravi2018a}
{Ravi} V.,  {Loeb} A.,  2018, arXiv e-prints, \href
  {http://adsabs.harvard.edu/abs/2018arXiv181100109R} {}

\bibitem[\protect\citeauthoryear{{Ravi} et~al.,}{{Ravi}
  et~al.}{2019}]{ravi2019}
{Ravi} V.,  et~al., 2019, arXiv e-prints, \href
  {https://ui.adsabs.harvard.edu/abs/2019arXiv190701542R} {p. arXiv:1907.01542}

\bibitem[\protect\citeauthoryear{{Remazeilles}, {Dickinson}, {Banday},
  {Bigot-Sazy}  \& {Ghosh}}{{Remazeilles} et~al.}{2015}]{rema2015}
{Remazeilles} M.,  {Dickinson} C.,  {Banday} A.~J.,  {Bigot-Sazy} M.~A.,
  {Ghosh} T.,  2015, \mn@doi [\mnras] {10.1093/mnras/stv1274}, \href
  {https://ui.adsabs.harvard.edu/abs/2015MNRAS.451.4311R} {451, 4311}

\bibitem[\protect\citeauthoryear{Rousseeuw \& Croux}{Rousseeuw \&
  Croux}{1993}]{Rousseeuw1993}
Rousseeuw P.~J.,  Croux C.,  1993, \mn@doi [Journal of the American Statistical
  Association] {10.1080/01621459.1993.10476408}, 88, 1273

\bibitem[\protect\citeauthoryear{{Sankrit} et~al.,}{{Sankrit}
  et~al.}{1998}]{san98}
{Sankrit} R.,  et~al., 1998, \apj, 504, 344

\bibitem[\protect\citeauthoryear{Shannon et~al.,}{Shannon
  et~al.}{2018}]{sh2018}
Shannon R.~M.,  et~al., 2018, \mn@doi [Nature] {10.1038/s41586-018-0588-y},
  562, 386

\bibitem[\protect\citeauthoryear{{Sokolowski} et~al.,}{{Sokolowski}
  et~al.}{2018}]{sok2018}
{Sokolowski} M.,  et~al., 2018, \mn@doi [\apjl] {10.3847/2041-8213/aae58d},
  \href {http://adsabs.harvard.edu/abs/2018ApJ...867L..12S} {867, L12}

\bibitem[\protect\citeauthoryear{{Spitler} et~al.,}{{Spitler}
  et~al.}{2016}]{sp16}
{Spitler} L.~G.,  et~al., 2016, Nature, 531, 202

\bibitem[\protect\citeauthoryear{{Syunyaev}}{{Syunyaev}}{1971}]{sun1971}
{Syunyaev} R.~A.,  1971, \azh, \href
  {https://ui.adsabs.harvard.edu/abs/1971AZh....48..244S} {48, 244}

\bibitem[\protect\citeauthoryear{{The CHIME/FRB Collaboration} et~al.,}{{The
  CHIME/FRB Collaboration} et~al.}{2019}]{CHIME2019c}
{The CHIME/FRB Collaboration} et~al., 2019, arXiv e-prints, \href
  {https://ui.adsabs.harvard.edu/abs/2019arXiv190803507T} {}

\bibitem[\protect\citeauthoryear{{Thornton} et~al.,}{{Thornton}
  et~al.}{2013}]{th13}
{Thornton} D.,  et~al., 2013, Science, 341, 53

\bibitem[\protect\citeauthoryear{Yao, Manchester  \& Wang}{Yao
  et~al.}{2017}]{Yao2017}
Yao J.~M.,  Manchester R.~N.,   Wang N.,  2017, \mn@doi [The Astrophysical
  Journal] {10.3847/1538-4357/835/1/29}, 835, 29

\bibitem[\protect\citeauthoryear{{van Haarlem} et~al.,}{{van Haarlem}
  et~al.}{2013}]{vanHaarlem2013}
{van Haarlem} M.~P.,  et~al., 2013, \mn@doi [\aap]
  {10.1051/0004-6361/201220873}, \href
  {https://ui.adsabs.harvard.edu/abs/2013A%26A...556A...2V} {556, A2}

\bibitem[\protect\citeauthoryear{{van Straten} \& {Bailes}}{{van Straten} \&
  {Bailes}}{2011}]{vanStraten2012}
{van Straten} W.,  {Bailes} M.,  2011, \mn@doi [\pasa] {10.1071/AS10021}, \href
  {https://ui.adsabs.harvard.edu/abs/2011PASA...28....1V} {28, 1}

\makeatother
\end{thebibliography}








\bsp	
\label{lastpage}
\end{document}